\documentclass[a4paper,onecolumn,oneside,11pt]{article}
\usepackage{geometry,color}
\usepackage{amsmath, amsfonts,amssymb,theorem,
euscript,array,enumerate,amsfonts,mathrsfs,xcolor}
\usepackage[dvips]{graphicx}
\usepackage[dvips]{graphics}
\usepackage{epstopdf}
\usepackage[dvips]{epsfig}
\input xy
\xyoption{all}

\definecolor{Ying}{rgb}{0.8,0,0.5}

\newcommand{\esp}{\mathbb{E}} 

\newcommand{\indic}{1\hspace{-2.7pt}\mathrm{l}} 
\newcommand{\proba}{\mathbb{P}}

\newcommand{\bF}{\mathbb{F}}

\newcommand{\bG}{\mathbb{G}}
\newcommand{\bH}{\mathbb{H}}

\newcommand{\N}{\mathcal{N}}

\newcommand{\bde}{\begin{displaymath}}
\newcommand{\ede}{\end{displaymath}}
\newcommand{\el}{\end{lem}}
\newcommand{\be}{\begin{equation}}
\newcommand{\ee}{\end{equation}}
\newcommand{\beq}{\begin{eqnarray*}}
\newcommand{\eeq}{\end{eqnarray*}}
\newcommand{\beqa}{\begin{eqnarray}}
\newcommand{\eeqa}{\end{eqnarray}}
\newcommand{\bel }{\left\{\begin{array}{ll}}
\newcommand{\eel}{\cr \end{array} \right.}
\newcommand{\bex}{\begin{ex} \rm }
\newcommand{\eex}{\end{ex}}

\newcommand{\bp}{\begin{pro}}
\newcommand{\ep}{\end{pro}}

\newcommand{\rmi}{{\rm (i)$\>\>$}}
\newcommand{\rmii}{{\rm (ii)$\>\>$}}
\newcommand{\rmiii}{{\rm (iii)$\>\,    \,$}}

\newtheorem{pro}{Proposition}[section]

\def\F{{\cal F}}
\def\H{{\cal H}}
\def\G{{\cal G}}

\def \bH {{\mathbb{H}}}

\theoremstyle{plain}
\theorembodyfont{\upshape\it}
\newtheorem{Thm}{\bf Theorem}[section]
\newtheorem{Pro}[Thm]{\bf Proposition}
\newtheorem{Lem}[Thm]{\bf Lemma}
\newtheorem{Cor}[Thm]{\bf Corollary}
\theorembodyfont{\upshape}

\newtheorem{Exe}[Thm]{Example}
\newtheorem{Rem}[Thm]{Remark}

\newtheorem{Hyp}[Thm]{Hypothesis}

\newcommand {\proof} {\noindent {\sc Proof:  }}
\newcommand {\finproof} {\hfill $\Box$ \vskip 5 pt }

\def\edoc{\end{document} }
\def\lbr{[\![}
\def\rbr{]\!]}
\newcommand{\bsigma}{\boldsymbol{\sigma}}

\newcommand{\btau}{\boldsymbol{\tau}}

\newcommand{\bu}{\boldsymbol{u}}

\newcommand{\bL}{\boldsymbol{L}}

\begin{document}
\thispagestyle{empty}
\title{Dynamics of multivariate default system in random environment\thanks{We thank Tahir Choulli, Delia Coculescu, Thierry Jeulin and Shiqi Song for interesting discussions. We are grateful to the two anonymous referees for helpful comments and suggestions. }}
\author{
Nicole El Karoui\thanks{Laboratoire de Probabilit\'es et Mod\`eles Al\'eatoires (LPMA), Universit\' e Pierre et Marie Curie - Paris 6, Paris, France; email: elkaroui@cmapx.polytechnique.fr} \quad
Monique Jeanblanc\thanks{Laboratoire de Math\'ematiques et Mod\'elisation d’\'Evry (LaMME), Universit\'e d'Evry-Val d'Essonne, UMR CNRS 8071, 91025 \'Evry Cedex France; email: monique.jeanblanc@univ-evry.fr}\quad Ying
Jiao\thanks {Institut de Science Financi\`ere et d'Assurances (ISFA), Universit\' e Claude Bernard - Lyon 1, 50 Avenue Tony Garnier, 69007 Lyon, France; email:
ying.jiao@univ-lyon1.fr} }
\date{\today}
\maketitle

\begin{abstract}We consider a multivariate default system where random environmental information is available. We study the dynamics of the system in a general setting of enlargement of filtrations and adopt the point of view of change of probability measures. We also make a link with the density approach in the credit risk modelling. Finally, we present a martingale characterization result with respect to the observable information filtration on the market. 

\vspace{2mm}

\vspace{1mm}
\end{abstract}

\section{Introduction }


We consider a system of finite default times to study their probability distributions and the dependence between the default system and the environmental market. 
 In the credit risk analysis, 
the environmental information appears to be an important factor. Besides the dependence structure among the underlying defaults, we also need to investigate the role of other market information upon the system of multiple defaults, and vice versa, the impact of default events on the market. In the credit risk modelling such as in the book of Bielecki and Rutkowski \cite{BR} and the paper of Elliott, Jeanblanc and Yor \cite{EJY}, the information structure concerning default times is described by the theory of enlargement of filtrations. In general, we suppose that on the market which is represented by a probability space $(\Omega,\mathcal A,\mathbb P)$, the environmental information is modelled by a reference filtration $\mathbb F=(\mathcal F_t)_{t\geq 0}$ and the default information is then added to form an enlarged filtration $\mathbb G=(\mathcal G_t)_{t\geq 0}$ which represents the global information of the market. The modelling of the dependence structure of multiple default times is then diversified in two directions by using the bottom-up and top-down models. In the former approach, one starts with a model for the marginal distribution of each default time and then the correlation between them is made precise (see Frey and McNeil \cite{FM2001} for a survey).  While in the top-down models which are particularly developed for the portfolio credit derivatives (see for example Arnsdorff and
Halperin \cite{AH2008},  Bielecki, Cr\'epey and Jeanblanc \cite{BCJ}, Cont and Minca \cite{CM}, Dassios and Zhao \cite{DZ2011}, Ehlers and Sch\"onbucher \cite{ES2009}, Filipovi\'c, Overbeck and Schmidt \cite{FOS}, Giesecke, Goldberg and Ding \cite{GGD}, Sidenius, Piterbarg and Andersen
\cite{SPA2008} among others), we study directly the cumulative loss process and its intensity dynamics. 

In this paper, we consider a multi-default system in presence of environmental information by using a general setting of enlargement of filtrations.  In order to fully investigate the different key elements in the modelling,  we use one random variable $\chi$ on $(\Omega,\mathcal A)$ valued in a polish space $E$ to describe default risks  and  to study the dependence between the default system and the remaining market. 
Given an observation filtration $(\mathcal N_t^E)_{t\geq 0}$ on  $(E,\mathcal E)$ with $\mathcal E=\mathcal B(E)$, the observable information associated to the default system $\chi$ is given by the inverse image filtration  $(\mathcal N_t:=\chi^{-1}(\mathcal N_t^E))_{t\geq 0}$ on $(\Omega,\mathcal A)$. The global market information observed at time $t\geq 0$ is then defined by the enlargement of $\mathcal F_t$ as $\mathcal G_t=\mathcal F_t\vee\mathcal N_t$.  The main advantage of this presentation is twofold. First, the  setting is general and can be applied flexibly to diverse situations by suitably choosing the default variable $\chi$ and the observation filtration. For example, for a multi-default system, we can include both ordered (which corresponds to top-down models) and non-ordered (which corresponds to bottom-up models) default times. Second, this framework allows us to distinguish the dependence structures of different nature, notably the correlation within the default system characterized by the so-called prediction process and the dependence between the default system and the environmental market described by a change of probability measure. 

The prediction process has initially been  introduced in the reliability theory (see for example Norros \cite{Nor}  and Knight \cite{Kn1975}) which is defined as the conditional law of $\chi$ with respect to its observation history $\mathcal N_t$ and describes the  dynamics of the whole default system upon each default event. 
When we take into account the environmental information $\mathcal F_t$, the market information is represented by an enlarged filtration. The main idea is to characterize the dependence between the multi-default system and the remaining market by using a change of probability method with respect to the product probability measure under which the multi-default system $\chi$ and the environmental information $\mathcal F_t$ are independent.  In this setting, the dependence structure between the default system and the market environment under any arbitrary probability measure can be described in a dynamic manner and represented by the Radon-Nikodym derivative process of the change of probability. We begin our analysis by focusing on the product probability space. In this case,  it is easy to apply the change of probability method to deduce estimation and evaluation formulas with market information.  
The general case where the market is not necessarily represented by a product probability space is more delicate, especially when the Radon-Nikodym derivative is not supposed to be strictly positive as usual. But the previous special case of product space will serve as a useful tool. We show that the key elements in the computation are indeed the prediction process and the Radon-Nikodym derivative.

We establish a link of the change of probability method with the default density approach in El Karoui, Jeanblanc and Jiao \cite{ejj1, ejj2}. In classic literature on enlargement of filtration theory, the density hypothesis was first introduced by Jacod \cite{Ja1987} in an initial enlargement of filtration and is fundamental to ensure the semi-martingale property in the enlarged filtration. We show that the density process and the Radon-Nikodym derivative can be deduced from each other and the two approaches are closed related. Finally, we are interested in the martingale processes in the market filtration $\mathbb G$. We present a general martingale characterization result which can be applied to  cases such as ordered and non-ordered defaults, which  are useful for financial applications. We also discuss the immersion property and show how to construct a $\mathbb G$-martingale by using the  characterization result.

The following of the paper is organized as follows. In Section 2, we give the general presentation of the multivariate default system and the market filtration. Section 3 focuses on the interaction between the default system and environmental information and presents the change of probability measure methodology in the product probability space setting.  In Section 4, we consider the general setting and investigate the link to the density approach in the theory of enlargement of filtrations. Finally in Section 5, we present the martingale characterization results. 

\section{The multi-default system}\label{section 1 default}

In this presentation, we introduce 
a general variable $\chi$ to  describe all uncertainty related to the multi-default system such as default or failure times, occurrence orders and   associated losses or recovery  ratios. We present different levels of information on the market and in particular a general construction of observable information filtration.

\subsection{Model setting}
\label{subsec: setup}

We fix a probability space $(\Omega,\mathcal A,\mathbb P)$. Let us consider a finite family of $n$ underlying firms  and describe all default uncertainty by a random variable $\chi$ which  takes values in a Polish space $E$. 
The default times of these firms are represented by a vector of random times $\btau=(\tau_1,\cdots,\tau_n)$. Since $\chi$ contains all information about default uncertainty, there exists a measurable map $f:E\rightarrow\mathbb R_+^n$ such that $\btau=f(\chi)$ and this map $f$ {specifies} the default times $\btau$.  In this general framework, the random variable $\chi$ can be chosen in a very flexible manner, which allows to consider bottom-up and top-down models in the credit risk literature. For example, one can choose $\chi$ to be the default time vector $\btau$ itself: in this case the Polish space $E$ is just $\mathbb R_+^n$ and $f:E\rightarrow\mathbb R_+^n$ is the identity map. One can also take into account the information of associated losses, namely $E=\mathbb R_+^n\times\mathbb R^n$ and $\chi=(\tau_i,L_i)_{i=1}^n:\Omega\rightarrow\mathbb R_+^n\times\mathbb R^n$, where $L_i$ denotes the loss induced by the $i^{\text{th}}$ firm at default time. In the top-down models, we consider the ordered default times $\sigma_1\leq\cdots\leq\sigma_n$. We can choose $E$ to be the subspace 
$\{\bu=(u_1,\cdots,u_n)\in\mathbb R_+^n\,|\,u_1\leq\cdots\leq u_n\}$ of $\mathbb R_+^n$
and $\chi$ to be the successive default vector $\bsigma=(\sigma_1,\cdots,\sigma_n)$. If we intend to take into account the label of each defaulted firm in the top-down setting (in this case, the successive default vector $\bsigma$ consists of the order statistics of the random vector $\btau$), we can choose $\chi=(\bsigma,J)$ valued in 
$E=\{\bu\in\mathbb R_+^n\,|\,u_1\leq \cdots\leq u_n\}\times\mathfrak S_n,$
where  $J$ takes values in the permutation group $\mathfrak S_n$  of all bijections from $\{1,\ldots,n\}$ to itself and describes the indices of underlying components for the successive defaults. The default time vector $\btau$ can thus be specified by the measurable map from $E$ to $\mathbb R_+^n$ which sends $(u_1,\ldots,u_n,\pi)$ to $(u_{\pi^{-1}(1)},\cdots,u_{\pi^{-1}(n)})$.

\subsection{Different information levels and filtrations}\label{subsec: filtrations}
We next present different levels of information: the total information and the accessible information.  Let $\mathbb F=(\mathcal F_t)_{t\geq 0}$ be a filtration of $\mathcal A$ and represent the environmental information of the market which is not directly related to the default events. We suppose that $\mathbb F$ satisfies the usual conditions and that $\mathcal F_0$ is the trivial $\sigma$-algebra. The observable default information on $E$ is described by a filtration of the $\sigma$-algebra $\sigma(\chi)$ which we precise below. 

The first
natural filtration associated {with} this framework is the
regularization of the filtration $$\bH=(\H_t)_{t\geq 0} \,\text{ where }\, \mathcal H_t:=\F_t\vee\sigma(\chi)$$ 
%
%
The $\sigma$-algebra $\H_t$ represents the total information  and is not completely observable since the default variable $\chi$ is not known by investors at an arbitrary time $t$. 
In the literature of enlargement of filtration, this filtration is called the initial enlargement of $\bF$ by $\chi$.  We recall that (e.g. Jeulin \cite[Lemma 4.4]{Je1980}, see also Song \cite{Song}) any $\mathcal H_t$-measurable random variable can be written in the form $Y_t(\chi)$ where $Y_t(\cdot)$ is an $\mathcal F_t\otimes\mathcal E$-measurable function.

The total information on $\chi$ is not accessible to all market participants at time $t\geq 0$.
We represent the  observable information flow  on the default events by a filtration $(\mathcal N_t)_{t\geq 0}$  on $\mathcal A$.  Typically it can be chosen as the 
filtration generated by a counting process. 
In this paper, we construct  the  filtration $(\mathcal N_t)_{t\geq 0}$ in a more general way. Let $(\mathcal N_t^E)_{t\geq 0}$ be a 
filtration on the Borel $\sigma$-algebra $\mathcal E=\mathcal B(E)$. If  $(\mathcal N^E_t)_{t\geq 0}$  is generated by an observation process $(N_t, t\geq 0)$ such as the default counting process or the cumulative loss process defined on the Polish space $E$, then  the filtration $(\mathcal N_t)_{t\geq 0}$ on $\Omega$ is defined as the inverse image (see e.g. Resnick \cite[\S 3.1]{Res99}) which is generated by the process $(N_t\circ\chi,t\geq 0)$.
More generally, the filtration $(\mathcal N_t^E)_{t\geq 0}$ determines a filtration on $(\Omega, \mathcal A)$ by the inverse image  by $\chi$   as 
\begin{equation}\label{inverse image}\mathcal N_t:=\chi^{-1}(\mathcal N_t^E)=\{\chi^{-1}(A)\,|\,A\in\mathcal N_t^E\}, \quad t\geq 0.\end{equation}
The global observable market information is then given by the filtration $\mathbb G=(\G_t)_{t\geq 0}$ where $$\G_t=\cap_{s>t} \,(\F_s\vee \N_s), \quad t\geq 0.$$  Note that $\mathbb G$ satisfies the usual conditions, and the following relation holds
\[\mathcal F_t\subseteq\mathcal G_t\subseteq\mathcal H_t.\] 


The filtration $\mathbb G$ can be viewed as a progressive enlargement but built in a general way. 
The classic initial and progressive enlargements are included in this framework. Below are some examples.
\begin{Exe} \label{example 1} \begin{enumerate}[(1)]
 \item If $\mathcal N_t^E=\mathcal E$ for all $t$, then $\mathcal N_t=\chi^{-1}(\mathcal E)=\sigma(\chi)$. So the filtration $\mathbb G$ coincides with $\mathbb H$, which is  the initial enlargement of filtration $\mathbb F$ by $\chi$.

\item In the case $n=1$  and $E=\mathbb R_+$ where $\chi=\tau$ denotes the default time of a single firm, for any $t\geq 0$, let $\mathcal N^E_t$ be generated by the functions of the form $\indic_{[0,s]}$ with $s\leq t$, 
then $\mathcal N_t=\sigma(\indic_{\{\tau\leq s\}},s\leq t)$. The filtration $\mathbb G$ is the classic progressive enlargement of $\mathbb F$ by $\tau$.

\item In general, $(\mathcal N_t^E)_{t\geq 0}$ can be any filtration on $(E,\mathcal E)$ so that  $\mathbb G$ differs from the initial and progressive enlargements.  Consider an insider who has extra information and knows if the firm will default before a deterministic time $t_0$ or not. Let $\chi=\tau$ denote the default time. For any $t\geq 0$, the $\sigma$-algebra $\mathcal N^E_t$ is  generated by the functions of the form $\indic_{[0,s]}$, with $s\leq t$ or $s=t_0$. Then the filtration of the insider is given by 
$\mathcal N_t=\tau^{-1}(\mathcal N_t^E)=\sigma(\indic_{\{\tau\leq s\}},\;s\leq t\text{ or }s=t_0)$, $t\geq 0.$
The filtration $\mathbb G=(\mathcal F_t\vee\mathcal N_t)_{t\geq 0}$ is in general larger than the progressive enlargement of $\mathbb F$ by $\tau$.

\item  
In the case with advanced or delayed information, if an agent knows the default occurrence with time $\epsilon>0$ in advance, then $\mathcal N^E_t$ is generated by the functions of the form $\indic_{[0,s+\epsilon]}$ with $s\leq t$ and  $\mathcal N_t=\sigma(\indic_{\{\tau\leq s+\epsilon\}},s\leq t)$. Similarly, for the delayed information  as in Collin-Dufresne, Goldstein and Helwege \cite{CDGH} or Guo, Jarrow and Zeng \cite{GJZ},  $\mathcal N^E_t$ is generated by $\indic_{[0,(s-\epsilon)^+]}$ with $s\leq t$ and  $\mathcal N_t=\sigma(\indic_{\{\tau\leq (s-\epsilon)^+\}},s\leq t)$. 
\item For a family of default times of $n$ non-ordered firms where $\chi=\btau$, we have $E=\mathbb R_+^n$. Let $\mathcal N^E_t$ be generated by the family of functions $\{\indic_{\{u_i\in[0,s]\}}(\bu), i\in\{1,\cdots, n\}, s\leq t\}$ with $\bu=(u_1,\cdots, u_n)$. Then $\mathcal N_t=\sigma(\indic_{\{\btau\leq s\}},s\leq t)$ describes the occurrence of default events. 
If the default times are ordered, we let $\chi=\bsigma$ and $E=\{(u_1,\cdots,u_n)\in\mathbb R_+^n\,|\,u_1\leq\cdots\leq u_n\}$. In this case, instead of including each default indicator, the counting process gives the information of successive defaults (see \cite{ejj2}). Let
$\mathcal N^E_t$ be generated by the functions of the form $\sum_{i=1}^n\indic_{\{u_i\in[0,s]\}}(\bu)$ with $s\leq t$. Then  $\mathcal N_t=\sigma(\indic_{\{\bsigma\leq s\}},s\leq t)$. 

\item If, besides the default occurrence, we also include a non-zero random mark $L_i$ at each default time such as the loss or recovery given default, we let $\chi=(\btau, \bL)$, which takes value in $E=\mathbb R_+^n\times\mathbb R^n$. 
For any $t\geq 0$, let $\mathcal N_t^E$ be generated by the functions of the form
$\{\indic_{\{u_i\in[0,s]\}}(\bu)f(l_i): i\in\{1,\cdots, n\},\; s\leq t,\;f\text{ is any Borel function on $\mathbb R$}\}$. Then the corresponding default filtration on $(\Omega,\mathcal A)$ is given by $\mathcal N_t=\sigma(\indic_{\{\tau_i\leq s\}}L_i,i\in\{1,\cdots, n\},\;s\leq t)$. 
Similarly, if the default times are ordered and $\chi=(\bsigma, \bL)$, then we can consider the marked point process (see e.g. Jacod \cite{Ja75}). The $\sigma$-algebra  $\mathcal N_t^E$ is generated by the functions of the form
$\sum_{i=1}^n\indic_{\{u_i\in[0,s]\}}(\bu)f(l_i), s\leq t$. Then 
$\mathcal N_t=\sigma\big(\sum_{i=1}^n\indic_{\{\sigma_i\leq s\}}L_i,\;s\leq t\big)$ as the random marks $L_i$ do not take the zero value.
\end{enumerate} 
\end{Exe}

\section{Prediction process and product space}

We are interested in the conditional law of $\chi$ under different information. In this section, we first focus on a situation which only concerns the default information and we introduce the notion of prediction process following Knight \cite{Kn1975} and Norros \cite{Nor}. We then include the environmental information $\mathbb F$ in a toy model of product probability space and show that the prediction process plays a key role in this case.    

\subsection{Prediction process}\label{Subsec:Prediction}
Given the default observation information,  the $(\mathcal N_t)_{t\geq 0}$-conditional probability law of the random variable $\chi$ is given as a $\mathcal P(E)$-valued c\`adl\`ag $(\mathcal N_t)_{t\geq 0}$-adapted process $ (\eta_t, t\geq 0)$, where $\mathcal P(E)$ denotes the set of all Borel probability measures on $E$, equipped with the topology of weak convergence  such that for any $\nu\in\mathcal P(E)$ and  any bounded continuous function $h$ on $E$, the map $\nu\longmapsto \int_E h\,d\nu$ is continuous. 

Using the terminology in \cite{Kn1975, Nor}, we call the process $(\eta_t, t\geq 0)$  the \emph{prediction process} of the random variable $\chi$ with respect to the observation filtration $(\mathcal N_t)_{t\geq 0}$. We refer the reader to \cite[Theorem 1.1]{Nor}  for the existence of a c\`adl\`ag version of the process $(\eta_t, t\geq 0)$ and the uniqueness up to indistinguishability. Moreover, the process $(\eta_t,t\geq 0)$ is an $(\mathcal N_t)_{t\geq 0}$-martingale with respect to the weak topology in the following sense~: for any bounded Borel function $h$ on $E$, the integral process $(\int_E h\,d\eta_t,t\geq 0)$ is an $(\mathcal N_t)_{t\geq 0}$-martingale.

\vspace{-3mm}
\paragraph{Example of a default counting process with random marks.} {We consider the case where the observation filtration $(\mathcal N_t^E)_{t\geq 0}$ is generated by the marked point process \[\sum_{i=1}^n\indic_{\{u_i\leq t\}}f(l_i),\;t\geq 0\]
which is associated to the occurrence sequence of default events together with the random marks as in Example \ref{example 1} (6). Note that the corresponding  $\mathcal N_t$ defined in \eqref{inverse image} identifies with the $\sigma$-algebra generated by the vector $(\bsigma,\bL)_{(i)}:=(\sigma_k,L_k)_{k=1}^i$ on the set $\{N_t^{\bsigma}=i\}=\{\sigma_i\leq t<\sigma_{i+1}\}$, where $N_t^{\bsigma}=\sum_{i=1}^n\indic_{\{\sigma_i\leq t\}}$.} Moreover, as we have mentioned, the default vector $\bsigma=(\sigma_1,\ldots,\sigma_n)$ can be written in the form $\bsigma=(u_1(\chi),\ldots,u_n(\chi))$ where $u_1,\ldots, u_n$ are measurable functions on $E$.

The prediction process $(\eta_t,t\geq 0)$  at time $t\geq 0$, i.e. 
$\eta_t(dx)=\mathbb P(\chi\in dx\,|\,\mathcal N_t)$ can be calculated by using the Bayesian formula and  taking into consideration each event $\{\sigma_i\leq t<\sigma_{i+1}\}$ on which $\eta_t$ is obtained as the conditional distribution  of $\chi$ given $(\bsigma,\bL)_{(i)}$ restricted on the survival set $\{\sigma_{i+1}>t\}$ and normalized by the conditional survival probability of $\sigma_{i+1}$ given $(\bsigma,\bL)_{(i)}$, that is, 
\begin{equation}\label{mu}
\eta_t(dx)=\sum_{i=0}^{n}\indic_{\{\sigma_i\leq t<\sigma_{i+1}\}}\frac{\eta_{|(\bsigma,\bL)_{(i)}}(\indic_{\{t<u_{i+1}(x)\}}\cdot dx)}{\eta_{|(\bsigma,\bL)_{(i)}}(\indic_{\{t<u_{i+1}(\cdot)\}})}=\frac{\eta_{{|}(\bsigma,\bL)_{(N_t)}}(\indic_{\{t<u_{N_t+1}(x)\}}\cdot dx)}{\eta_{{|}(\bsigma,\bL)_{(N_t)}}(\indic_{\{t<u_{N_t+1}(\cdot)\}})},
\end{equation}
where $\eta_{|(\bsigma,\bL)_{(i)}}$ is the conditional law of $\chi$ given $(\bsigma,\bL)_{(i)}$, and $\eta_{|(\bsigma,\bL)_{(i)}}(\indic_{\{t<u_{i+1}(x)\}}\cdot dx)$ denotes the random measure on $E$ sending a bounded Borel function $h:E\rightarrow\mathbb R$ to
\[\int_{E}h(x)\,\eta_{|(\bsigma,\bL)_{(i)}}(\indic_{\{t<u_{i+1}(x)\}}\cdot dx):=\mathbb E[h(\chi)\indic_{\{t<\sigma_{i+1}\}}\,|\,(\bsigma,\bL)_{(i)}].\] At each default time,  the new arriving default event brings a regime switching to the  prediction process, which can be interpreted as the impact of  default contagion phenomenon. 

In the particular case where $\chi$ coincides with $(\bsigma,\bL)$ and  the probability law of $(\bsigma,\bL)$ has a density $\alpha(\cdot,\cdot)$ with respect to the Lebesgue measure, similar as in \cite{ejj2}, we obtain a more explicit form of the prediction process as follows~:
\[\eta_t(d(\boldsymbol{v},\boldsymbol{l}))=\sum_{i=0}^{n}\indic_{\{\sigma_i\leq t<\sigma_{i+1}\}}\delta_{(\bsigma,\bL)_{(i)}}(d(\boldsymbol{v},\boldsymbol{l})_{(i)})\indic_{\{t<v_{i+1}\}}\frac{\alpha(\boldsymbol{v},\boldsymbol{l})\,d(\boldsymbol{v},\boldsymbol{l})_{(i+1:n)}}{\int_{t}^\infty\alpha(\boldsymbol{v},\boldsymbol{l})\,d(\boldsymbol{v},\boldsymbol{l})_{(i+1:n)}},\]
where $(\boldsymbol{v},\boldsymbol{l})_{(i)}=(v_k,l_k)_{k=1}^i$, $(\boldsymbol{v},\boldsymbol{l})_{(i+1:n)}=(v_{k},l_{k})_{k=i+1}^n$, and
\[\int_{t}^\infty\alpha(\boldsymbol{v},\boldsymbol{l})\,d(\boldsymbol{v},\boldsymbol{l})_{(i+1:n)}:=\int_{(]t,\infty[\,\times\mathbb R)^{n-i}}\alpha(\boldsymbol{v},\boldsymbol{l})\,d(v_{i+1},l_{i+1})\cdots d(v_n,l_n).\]

\subsection{Conditional default distributions under the product measure}\label{subsection: product space}

We now consider the conditional law of $\chi$ given the general observation information $\mathbb G$ but in a simple case of product space.
We assume that the global market $(\Omega,\mathcal A)$ can be written as $(\Omega^\circ\times E,\mathcal A^\circ\otimes\mathcal E)$ and that the filtration $\mathbb F$ is given by $(\mathcal F_t^{\circ}\otimes\{\emptyset,E\})_{t\geq 0}$, where $(\Omega^\circ,\mathcal A^\circ)$ is a measurable space and $\mathbb F^\circ=(\mathcal F_t^\circ)_{t\geq 0}$ is a filtration of $\mathcal A^\circ$. 
The $\mathcal A$-measurable random variable $\chi$ is assumed to be given by the second projection, i.e. $\chi(\omega,x)=x$. The filtration $\mathbb H$ of total information is then $(\mathcal H_t=\mathcal F_t^\circ\otimes\mathcal E)_{t\geq 0}$. We let $\mathbb P^\circ$ be the marginal probability measure of $\mathbb P$ on $(\Omega^\circ,\mathcal A^\circ)$, namely for any bounded $\mathcal A^\circ$-measurable $f$ on $\Omega^\circ$, 
$\int_{\Omega^\circ}f(\omega^\circ)\mathbb P^\circ(d\omega^\circ):=\int_{\Omega}f(\omega^\circ)\mathbb P(d\omega^\circ,dx).$
Let $\eta$ be the probability law of $\chi$. In the product space, the probability measure $\eta$ identifies with the marginal measure of $\mathbb P$ on $(E,\mathcal E)$. 

The most simple dependence structure between  the default variable $\chi$ and the environmental information $\mathbb F$ is when they are independent. We introduce the product probability measure $\overline{\proba}=\mathbb P^\circ\otimes\eta$ on  $(\Omega,\mathcal A)$ under which the two sources of risks $\chi$ and $\mathbb F$ are independent. This case will serve as the building stone in our paper. In particular, the calculations are easy applications of Fubini's theorem in this case. 

We fix some notation which will be useful in the sequel. If $Y$ is a random variable on the product space $\Omega=\Omega^\circ\times E$, sometimes we omit the first coordinate in the expression of the $\mathcal A$-measurable function $Y$ and use the notation $Y(x)$, $x\in E$,  which denotes in fact the random variable $Y(\cdot,x)$.
\vspace{-3mm}
\paragraph{Default information.}  Recall that $\eta_t$ is the conditional probability law of $\chi$ given $\mathcal N_t$ and $(\eta_t, t\geq 0)$ is the prediction process of $\chi$. Since $\chi$ is independent of $\mathbb F$ under $\overline{\mathbb P}$,  for any bounded or positive  $\mathcal A$-measurable function $\Psi$ on $\Omega=\Omega^\circ\times E$, one has \[\esp_{\overline{\proba}}[\Psi\,|\,\F_{\infty}\vee \N_t]=\int_E \Psi(\cdot,x) \eta_t(dx)=:\eta_t(\Psi), \quad \overline{\mathbb P}\text{-}a.s.\]
By Dellacherie and Meyer \cite[VI.4]{DM2}, there exists a c\`adl\`ag version of the martingale $(\eta_t(\Psi),t\geq 0)$ as conditional expectations.

\paragraph{Total information $\mathbb H$. }For the case of total information $\mathbb H$,  we have to take care about negligible sets. In full generality, the equality $X(\cdot, x)=Y(\cdot,x)$, $\mathbb P^\circ$-a.s. for all $x\in E$ does not imply $X(\cdot,\chi)=Y(\cdot,\chi)$, $\overline{\mathbb P}$-a.s.. We need a suitable version for such processes. This difficulty can be overcome by Meyer \cite{Meyer} and Stricker and Yor \cite{SY1978}.

\begin{enumerate}[(i)]
\item  Given a non-negative $\mathcal A$-measurable function $\Psi$ on $\Omega$, from \cite{Meyer, SY1978}, there exists a c\`adl\`ag $\bH$-adapted process 
$(\Psi^\F_t(\cdot),t\geq 0)$ such that, for any $x \in E$, and for any $t\geq 0$, $$\Psi^\F_t(x)=\esp_{\proba^\circ}[\Psi(\cdot,x)|\F^{\circ}_t], \quad \mathbb P^\circ\text{-}a.s. $$ In particular, if $X_t$ is an  $\F^{\circ}_t$-measurable random variable valued in $E$, then one has 
\[\Psi^\F_t(X_t)= \esp_{\proba^{\circ}}[\Psi(\cdot,X_t)|\F^{\circ}_t],\quad \proba^\circ\text{-}a.s. \]
We call $(\Psi_t^\F(x),t\geq 0)_{x\in E}$  a {parametrized $({\mathbb F}^{\circ},{\mathbb P}^{\circ})$-martingale} depending on a parameter $x\in E$, since the conditional expectation property is valid for all values of $x$ and $t$ outside of a null set.
\item This parametrized version of  $\mathbb F^{\circ}$-conditional expectation  as a function of both variables $(\omega, x)$ is  the basic tool for studying projections with respect to $\mathbb H$. Under the product measure $\overline{\proba}=\mathbb P^\circ\otimes \eta$,
\begin{equation}\label{universal martingale}
\esp_{\overline\proba}[\Psi|\H_t]=\Psi^\F_t(\chi), \quad\overline{\proba}\text{-a.s.}\end{equation}
Furthermore, we can extend  $\eta_t$ to a $\mathcal G_t$-random measure on $(\Omega, \H_t)$ which sends any non-negative $\H_t$-measurable random variable $Y_t(\cdot)$ to the $\mathcal G_t$-measurable random variable $\eta_t(Y_t(\cdot))=\int_EY_t(x)\,\eta_t(dx)$. In the following, by abuse of language, we use $\eta_t$ to denote the conditional laws with respect to both $\mathcal N_t$ and $\mathcal G_t$ under $\overline{\mathbb P}$. 

\item The above result can be interpreted as a characterization of $(\bH,\overline{\proba})$-martingale in terms of a parametrized  $(\bF^{\circ},\proba^{\circ})$-martingale depending  on a parameter $x\in E$. We shall discuss the martingale properties in more detail in Section \ref{subsection: martingale}.
\end{enumerate}
\paragraph{Accessible information $\mathbb G$.}For the observable information $\mathbb G$, the projection is firstly made on a larger filtration which includes more information than $\mathbb G$ either on $\Omega^\circ$ or on $E$.\\
\rmi Since $\eta_t$ denotes  the conditional law of $\chi$ given both $\mathcal G_t$ and $\mathcal N_t$ under $\overline{\proba}$, for any non-negative $\H_t$-measurable random variable $Y_t(\chi)$, we have
\begin{equation}\label{mu G P0}
\esp_{\overline{\proba}}[Y_t(\chi)|\G_t]=\int_EY_t(x)\eta_t(dx)=\eta_t(Y_t(\cdot)).\end{equation}
\rmii Consider now a non-negative ${\mathcal A}$-measurable random variable $Y$ on ${\Omega}$. The calculation of its $\G_t$-conditional expectation   can be done in two different ways as shown below :
\begin{equation}\label{diagramme}
\xymatrix{\relax Y\ar[d]_{{\mathcal F}_\infty\vee\mathcal N_t}\ar[rr]^-{\mathcal H_t=\mathcal F_t\vee\sigma(\chi)}&&\mathbb E_{\overline{\proba}}[Y\,|\,\mathcal H_t]\ar[d]^-{\mathcal G_t}\\
\mathbb E_{\overline{\proba}}[Y\,|\,{\mathcal F}_\infty\vee\mathcal N_t]\ar[rr]_-{\mathcal G_t}&&\mathbb E_{\overline{\proba}}[Y\,|\,\mathcal G_t]}
\end{equation}
On the one hand, using the notation introduced in \eqref{universal martingale},
\[\esp_{\overline{\proba}}[Y|\G_t]=\esp_{\overline{\proba}}[\esp_{\overline{\proba}}[Y|\H_t]\,|\,\G_t]=\esp_{\overline{\proba}}[Y_t^{\F}(\chi)\,|\,\G_t]\]
which, by \eqref{mu G P0}, equals 
\begin{equation}\label{Y_T X}\esp_{\overline{\proba}}[Y|\G_t]
=\eta_t(Y_t^{\F}(\cdot))=\int_E Y_t^{\F}(x)\eta_t(dx).\end{equation}
On the other hand, as shown in \eqref{diagramme}, the above result can also be obtained by
using the intermediary $\sigma$-algebra $\F_\infty\vee\N_t$. Note that by the monotone class theorem, it suffices to consider $Y(\omega,x)$ of the form $ Y^{\circ}(\omega)h(x)$ where $Y^{\circ}$ is ${\F}^{\circ}_\infty$-measurable and $h$ is a Borel function on $E$, then
\[\begin{split}\esp_{\overline{\proba}}[Y|\G_t]&=\esp_{\overline{\proba}}[\esp_{\overline{\proba}}[Y^\circ h|\F_\infty\vee\N_t]\,|\,\G_t]
=\esp_{\overline{\proba}}[Y^\circ\eta_t(h)|\G_t]\\
&=\eta_t(h)\esp_{\mathbb P^{\circ}}[Y^\circ|\F^\circ_t]=\eta_t(\esp_{\mathbb P^{\circ}}[Y^\circ h|\F^\circ_t])=\eta_t(\esp_{\mathbb P^\circ}[Y|\F^\circ_t])\end{split}\]
which is equal to \eqref{Y_T X}.\\
\rmiii  By \eqref{Y_T X} we can characterize a $(\bG,\overline{\proba})$-martingale  as the integral of a parametrized $(\bF^\circ,\mathbb P^\circ)$-martingale $Y^{\F}(x)$, $x\in E$, with respect to the random measure $\eta_t(dx)$.

\subsection{Change of probability measures}\label{subsec:chgt proba product space}

In this subsection, we consider a probability measure $\proba$ on the product space $(\Omega, \mathcal A)=(\Omega^\circ\times E,\mathcal A^\circ\otimes\mathcal E)$ which is absolutely continuous with respect to the product measure $\overline{\mathbb P}$. We suppose that the Radon-Nikodym derivative of $\mathbb P$ with respect to the product measure $\overline{\proba}$ is given by 
\begin{equation}\label{chgt proba product space}
\frac{d\mathbb P}{d\overline{\proba}}\,\Big|_{\H_T}=\beta_{T}(\chi)
\end{equation}
where $T\geq 0$ is a horizon time and $\beta_T(\cdot)$ is a non-negative $\mathcal F_T\otimes\mathcal E$-measurable random variable. {We note that $\mathbb P$ is not necessarily an equivalent probability measure of $\overline{\mathbb P}$, that is, $\beta_T(\chi)$ is not supposed to be strictly positive $\overline{\mathbb P}$-almost surely.} The fact that $\eta$ identifies with the probability law of $\chi$ under the probability measure $\mathbb P$ implies that
\begin{equation}\label{condition beta}\esp_{\proba^{\circ}}[\beta_{T}(x)]=1,\quad \text{$\eta$-a.s. for $x\in E$}. \end{equation}
The dependence between the default variable $\chi$ and other market environment $\mathbb F$ under $\mathbb P$ is characterized by the change of probability in a dynamic manner. 

We still examine the different information levels. The conditional law of $\chi$ given $\N_t$ remains invariant under  $\proba$ and $\overline\proba$ because of \eqref{condition beta}, and is denoted as $\eta_t$ under both probability measures.\\[-10mm]
\paragraph{Total information $\mathbb H$. }The Radon-Nikodym density of $\mathbb P$ with respect to $\overline {\mathbb P}$ on $\H_t$ is specified by the parametrized $({\bF}^\circ,{\proba}^\circ)$-martingale $(\beta_t^{\F}(x)=\esp_{\mathbb P^\circ}[\beta_{T}(x)|\F^\circ_t],\,t\in[0,T])$ depending on the parameter $x\in E$. For any $t\in[0,T]$, $\beta_t^{\mathcal F}(\chi)$ is the Radon-Nikodym density of $\mathbb P$ with respect to $\overline{\mathbb P}$ on $\mathcal H_t$.
We use the notation  $\beta_t$ to denote $\beta_t^{\mathcal F}(\chi)$ when there is no ambiguity.  

\begin{enumerate}[(i)]
\item For a  non-negative $\mathcal H_T$-measurable random variable $Y$ on $\Omega$, by the change of probability and  Fubini's theorem under $\overline{\proba}$, we have
\begin{equation}\label{Y conditional H_t}\begin{split}
\esp_{\proba}[Y|\H_t]
=\frac{\esp_{\overline{\proba}}[Y\beta_{T}|\H_t]}{\beta_t}\indic_{\{\beta_t>0\}}
=\frac{(Y\beta)^{\F}_t}{\beta_t}\indic_{\{\beta_t>0\}},\quad \mathbb P\text{-}a.s.\end{split}\end{equation}
where the last equality comes from  \eqref{universal martingale}. 
\begin{Rem}\label{beta positive}
Note that we only suppose that $\mathbb P$ is absolutely continuous w.r.t. $\overline{\mathbb P}$. In particular, any $\overline{\mathbb P}$-negligible set is $\mathbb P$-negligible. However, the converse statement is not necessarily true. In particular, although $\beta_t$ is not necessarily strictly positive $\overline{\mathbb P}$-a.s., the set $\{\beta_t= 0\}$ is negligible under $\mathbb P$. In fact, one has
\[\mathbb P(\beta_t=0)=\mathbb E_{\overline{\mathbb P}}[\beta_t\indic_{\{\beta_t= 0\}}]=0\]
since $\beta_t$ is the Radon-Nikodym derivative $d\mathbb P/d\overline{\mathbb P}$ on $\mathcal H_t$ (note that this formula does not signify  $\overline{\mathbb P}(\beta_t=0)=0$). In the following, to simplify the presentation, we omit the indicator $\indic_{\{\beta_t>0\}}$ and the equality \eqref{Y conditional H_t} can be written as $\esp_{\proba}[Y|\H_t]=(Y\beta)_t^{\mathcal F}/\beta_t$. 

In addition, under the probability $\overline{\mathbb P}$, $\beta_T=0$ a.s. on the set $\{\beta_t=0\}$ 
since $\beta$ is a non-negative $(\mathbb H,\overline{\mathbb P})$-martingale. 
In fact, let $T^x=\inf\{t\geq 0:\beta_{t-}(x)=0\}$, then $\beta_t(x)>0$ on $\lbr 0, T^x\rbr$ and $\beta_t(x)=0$ on $\lbr T^x,\infty\rbr$.
\end{Rem}

\item By \eqref{Y conditional H_t},  an $\mathbb H$-adapted process  is an $(\bH,\proba)$-martingale if and only if its product with $\beta$ is an $(\bH,\overline{\proba})$-martingale, or equivalently, a parametrized $(\mathbb F^\circ, \mathbb P^\circ)$-martingale depending on a parameter $x\in E$.
\end{enumerate}
\paragraph{Accessible information $\mathbb G$.}
 We use the notation $\eta^{\G}$ for the conditional law of $\chi$ given $\G_t$ under the probability $\proba$. 

\begin{enumerate}[(i)] 
\item The Bayes formula allows to calculate directly the conditional law $\eta^{\G}$ by
\begin{equation}\label{Equ:mug}\eta^{\G}_t(dx)=\frac{\eta_t(\beta_t(x)\cdot dx)}{\eta_t(\beta_t(\cdot))}\end{equation}
where  for a non-negative $\mathcal A$-measurable function $\Psi$ on $\Omega$, the notation $\eta_t(\Psi(x)\cdot dx)$ denotes the $\mathcal A$-random measure on $E$ which sends a non-negative Borel function $f$ on $E$ to \begin{equation}\label{eta _t with function f}\int_E f(x)\eta_t(\Psi(x)\cdot dx)=\eta_t(f(\cdot)\Psi(\cdot))\end{equation}
\item The $\G_t$-conditional expectation of a non-negative $\H_t$-measurable random variable $Y_t(\chi)$ is 
\begin{equation}\label{random measure mu G}
\esp_{\proba}[Y_t(\chi)|\G_t]=\int_{E} Y_t(x)\eta^{\G}_t(dx)=:\eta^{\G}_t(Y_t(\cdot))\end{equation}
\item For a non-negative $\mathcal H_T$-measurable random variable $Y$ on ${\Omega}$, we first project  on the larger $\sigma$-algebra $\H_t$ and then use \eqref{Y conditional H_t} and \eqref{random measure mu G} to obtain
\begin{equation}\label{YTG}\esp_{\proba}[Y|\G_t]=\esp_{\proba}\big[\mathbb E_{\mathbb P}[Y|\H_t]|\G_t\big]
=\int_E \frac{(Y\beta)^{\F}_t(x)}{\beta_t(x)}\eta^{\G}_t(dx).
\end{equation}
An equivalent form can be obtained by using the Bayes formula  as 
\begin{equation}\label{Y T G2}\begin{split}\mathbb E_{{\mathbb P}}[Y\,|\,\mathcal G_t]
=\frac{\eta_t((Y\beta)^{\F}_t(\cdot))}{\eta_t(\beta_t(\cdot))}
\end{split}\end{equation}
The equality between \eqref{YTG} and \eqref{Y T G2} can also be shown by \eqref{Equ:mug}. 	
\item Accordingly, a $(\bG,\proba)$-martingale can be characterized as follows: a $\mathbb G$-adapted process is a $(\bG,\proba)$-martingale if and only if its product with $\eta_t(\beta_t(\cdot))$ is an integral of the parametrized  $(\mathbb F^\circ,\mathbb P^\circ)$-martingale $(Y\beta)^{\F}(x)$, $x\in E$ with respect to $\eta(dx)$, or alternatively, if and only if it can be written as the integral of the quotient of two parametrized $(\mathbb F^\circ,\mathbb P^\circ)$-martingales ${(Y\beta)^{\F}_t(x)}/{\beta_t(x)}$ with respect to  $\eta^{\G}_t(dx)$.
\end{enumerate}

\section{Interaction with the environment information} \label{section general case}

In this section, we come back to the general financial market setting modelled in Section \ref{subsec: setup}, where the probability space $(\Omega, \mathcal A, \mathbb P)$ is not necessarily given in a product form. However, the toy model of product space we discussed in the previous section will provide useful tools for the general case.  From the financial point of view, the method of change of probability measure allows to describe the dynamic dependence structure between the default risk variable $\chi$ and the underlying market information filtration $\mathbb F$. In a recent work, Coculescu \cite{Coc} has constructed an explicit non-Markovian default contagion model where the change of probability method is adopted. From the mathematical point of view, we obtain computational results where a standard hypothesis that the Radon-Nikodym derivative is supposed to be strictly positive can be relaxed.  

In the enlargement of filtrations, Jacod's density hypothesis (see \cite{Ja1987}), which is originally introduced in the setting of initial enlargement of filtrations,  plays an essential role. In the credit risk analysis, the density approach of default  has been adopted in progressive enlargement of filtrations to study the impact of a default event and a family of ordered multiple defaults (c.f. \cite{ejj1, ejj2}, Kchia, Larsson and Protter \cite{KLP} and Gapeev, Jeanblanc, Li and Rutkowski \cite{GJLR}). We present  the link between the conditional density and the Radon-Nikodym derivative  of change of probability in the general setting of a multi-default system. 


\subsection{Basic hypothesis}	

We first make precise the density hypothesis in our model setting. 
\begin{Hyp}\label{hyp:density} The conditional probability law of $\chi$ given the filtration $\mathbb F$ admits a density with respect to a $\sigma$-finite measure $\nu$ on $(E,\mathcal E)$, i.e., for any $t\geq 0$, there exists an $\mathcal F_t\otimes \mathcal E$-measurable function $(\omega,x)\rightarrow \alpha_t(\omega,x)$ such that for any non-negative Borel function $f$, \[\mathbb E_{\mathbb P}[f(\chi)|\mathcal F_t]=\int_Ef(x)\alpha_t(x)\nu(dx),\quad \mathbb P\text{-}a.s.\]
 In \cite{Ja1987}, $\nu$ is chosen to be $\eta$ which is the probability law of $\chi$. In practice, another choice of $\nu$ is often the Lebesgue measure, especially for the multi-default models. 
\end{Hyp}

Under the probability $\mathbb P$, for any $x\in E$, the process $(\alpha_t(x),t\geq 0)$ is an $\mathbb F$-martingale.  The probability law $\eta$ of $\chi$ is absolutely continuous with respect to the measure $\nu$ and is given by 
\[\eta(dx)=\alpha_0(x)\,\nu(dx).\]
In the case where $\nu$ coincides with $\eta$, we have $\alpha_0(x)=1$. The $\mathcal F_t$-conditional probability law  of $\chi$ is absolutely continuous with respect to $\eta$. Let
\begin{equation}\label{equ:beta def}\mathbb P(\chi\in dx|\mathcal F_t)=:\beta_t(x)\eta(dx),\end{equation}
then the following relation holds
\[\alpha_t(x)=\alpha_0(x)\beta_t(x),\quad\text{$\mathbb P\otimes\eta$-a.s.}.\]
By \cite[Lemma 1.8]{Ja1987}, there exists a non-negative c\`adl\`ag version of $(\omega,t,x)\mapsto\beta_t(\omega,x)$ such that for any $x\in E$, $\beta_t(x)$ is an $\mathbb F$-martingale.

We can interpret the density by using the language of change of probability measure.  In a setting of initial enlargement of filtrations, i.e. with respect to the filtration $\mathbb H$, where $\nu$ coincides with $\eta$, it is proved in Grorud and Pontier \cite{GP} (see also Amendinger, Becherer and Schweizer \cite{ABS}) that if  $\beta_t(\cdot)$ is strictly positive, $\mathbb P$-a.s., there exists a probability measure $\widehat{\mathbb P}$ on the measurable space $(\Omega,\mathcal A)$ which is equivalent to $\mathbb P$ such that  $\chi$ is independent of $\mathbb F$ under the probability measure $\widehat{\mathbb P}$, and that
$\widehat{\mathbb P}$ coincides with $\mathbb P$ on $\mathbb F$ and $\sigma(\chi)$ respectively. In fact, the process $(\frac{1}{\beta_t(\chi)}, t\geq 0)$ is a $(\mathbb H,\mathbb P)$-martingale of expectation $1$ and the probability $\widehat{\mathbb P}$ is characterized by the Radon-Nikodym derivative (c.f. \cite[Lemma 3.1]{GP})
\begin{equation}\label{proba hat}\frac{d\widehat{\mathbb P}}{d\mathbb P}\Big|_{\mathcal H_t}=\frac{\alpha_0(\chi)}{\alpha_t(\chi)}=\frac{1}{\beta_t(\chi)}.\end{equation}
The above result is very useful for studying insider's information. Inspired by this idea, we  study  processes in the observation filtration $\mathbb G$ (in the general setting which is not necessarily the initial or progressive enlargement of filtrations) by combining density and change of probability measure. 

\begin{Rem}\label{counter example}We note that the probability $\widehat{\mathbb P}$ does not exist in general on the probability space $(\Omega,\mathcal A)$, notably when $\eta_t^{\mathcal F}$ is absolutely continuous but not equivalent to $\eta$. In fact, although $\beta_t(\chi)>0$, $\mathbb P$-a.s. (see Remark \ref{beta positive}), in general, $\mathbb E_{\mathbb P}[\frac{1}{\beta_t(\chi)}]$ can be strictly smaller than $1$, so that we can't use \eqref{proba hat} to define an equivalent probability measure $\widehat{\mathbb P}$. 
We provide a simple counter-example. Let $\Omega=\{\omega_0,\omega_1\}$, with $\mathbb P(\{\omega_0\})=\mathbb P(\{\omega_1\})=\frac 12$ and $\mathcal F_0=\{\emptyset, \Omega\}$, $\mathcal F_1=\{\emptyset, \{\omega_0\}, \{\omega_1\}, \Omega\}$. Let $\chi:\Omega\rightarrow E=\{0,1\}$ be defined by $\chi(\omega_0)=0$ and $\chi(\omega_1)=1$. The $\mathcal F_1$-conditional law of $\chi$ is absolutely continuous with respect to $\eta$ with the density $\beta_1(\omega,x):\Omega\times E \rightarrow \mathbb R_+$ given by $2\indic_{\{\omega_0, 0\}}+2\indic_{\{\omega_1, 1\}}$. However, $\beta_1(\cdot)$ is not strictly positive on all senarios, and there does not exist a decoupling probability measure $\widehat{\mathbb P}$ under which $\chi$ is independent of $\mathbb F$ and that $\mathbb P$ is absolutely continuous with respect to $\widehat{\mathbb P}$. 
See Aksamit, Choulli and Jeanblanc \cite{ACJ} for a further discussion on this subject. 

In the general case where $\beta_t(\cdot)$ is not necessarily strictly positive, that is, the $\mathcal F_t$-conditional probability law  of $\chi$ is absolutely continuous but not equivalent with respect to its probability law $\eta$, we can no longer use the approach in \cite{GP} since the change of probability \eqref{proba hat} is not well defined. To overcome this difficulty, we propose to use a larger product measurable space constructed from  the initial probability space $(\Omega,\mathcal A)$. The results established in the previous section \ref{subsection: product space} will be useful, as we shall explain  in the next subsection. 
\end{Rem}


\subsection{Conditional expectations in the general setting}

This subsection focuses on the computations  with respect to the observable information $\mathbb G$ which is a general filtration as defined in Section \ref{subsec: filtrations}. 
Recall that $\mathbb G$ is an enlargement of $\mathbb F$ as $\mathcal G_t=\cap_{s>t} \,(\F_s\vee \N_s)$  where $(\mathcal N_t)_{t\geq 0}$ is the inverse image given by  $\mathcal N_t=\chi^{-1}(\mathcal N_t^E)$, and  satisfies the usual conditions. We still assume Hypothesis \ref{hyp:density}. Note that $\beta_t(\cdot)$ is not supposed to be strictly positive. 

To establish the main computation result, the idea in \cite{GP} is to use the decoupling probability measure where $\chi$ and $\mathbb F$ are independent. However, in the original space $(\Omega,\mathcal A)$, this probability measure $\widehat{\mathbb P}$ does not necessarily exist (see Remark \ref{counter example}).
Our method consists in extending the original probability space by introducing an auxiliary product space $(\Omega\times E,\mathcal A\otimes\mathcal E)$ which is equipped with a product probability measure $\overline{\mathbb P}=\mathbb P\otimes\eta$. We consider the graph map of $\chi$, which is by definition the map $\Gamma_\chi:\Omega\rightarrow\Omega\times E$ sending $\omega\in\Omega$ to $(\omega,\chi(\omega))$. Viewed as a random variable on $(\Omega,\mathcal A)$ valued in the product space $\Omega\times E$, the map $\Gamma_\chi$ admits a probability law   $\mathbb P'$ under $\mathbb P$. More precisely,  $\mathbb P'$ is the probability measure on the product space $(\Omega\times E,\mathcal A\otimes\mathcal E)$ such that,  for any non-negative $\mathcal A\otimes\mathcal E$-measurable function $f$ on $\Omega\times E$, 
\begin{equation}\label{equ:P'}\int_{\Omega\times E} f(\omega,x)\mathbb P'(d\omega,dx)=\int_{\Omega}(f\circ\Gamma_{\chi})(\omega)\mathbb P(d\omega)=\mathbb E_{\mathbb P}[f(\chi)]. \end{equation} 
Note that $\mathbb P'$ is absolutely continuous with respect to the product probability $\overline{\mathbb P}$, and the corresponding Radon-Nikodym derivative is given by $\beta_t(\cdot)$ on $\mathcal F_t\otimes\mathcal E$ under Hypothesis \ref{hyp:density} since
\[\mathbb E_{\mathbb P}[f(\chi)]=\mathbb E_{\mathbb P}[\mathbb E_{\mathbb P}[f(\chi)|\mathcal F_t]]=\int_\Omega\int_Ef(\omega,x)\beta_t(x)\eta(dx)\mathbb P(d\omega)=\int_{\Omega\times E}f(\omega,x)\beta_t(x)\overline{\mathbb P}(d\omega,dx)\]
if $f$ is $\mathcal F_t\otimes\mathcal E$-measurable. Moreover, the composition of $\Gamma_\chi$ with the second projection $\Omega\times E\rightarrow E$ coincides with  $\chi$. 
Thus we can use the method developed in Section \ref{subsec:chgt proba product space}. In the particular case where $\beta(\cdot)$ is strictly positive and the probability measure $\widehat{\mathbb P}$ in \eqref{proba hat} exists, then the probability law of $\Gamma_\chi$ under $\widehat{\mathbb P}$ coincides with the product probability measure $\overline{\mathbb P}$.

We first describe processes in the filtration $\mathbb G$. Note that if $Y(\cdot)$ is a function on $\Omega\times E$, then the expression $Y(\chi)$ denotes actually  $Y(\cdot)\circ\Gamma_{\chi}$ as  a function on $\Omega$.  We make precise in the following lemma the  measurability of the application: 
 \begin{equation}\label{equ:composition}\xymatrix{\relax \Omega\ar[r]^-{\Gamma_\chi}\ar@/_1.5pc/[rr]_-{Y(\chi)}&\Omega\times E\ar[r]^-{Y(\cdot)}&\mathbb R}.\end{equation} 
\begin{Lem}\label{Lem:measurability} 
 Let $\mathcal F$ be a sub-$\sigma$-algebra of $\mathcal A$ on $\Omega$ and $\mathcal E_0$ be a sub-$\sigma$-algebra of $\mathcal E$ on $E$. Then
\begin{enumerate}[1)]
\item the map $\Gamma_{\chi} : (\Omega,\mathcal F\vee\chi^{-1}(\mathcal E_0))\rightarrow (\Omega\times E, \mathcal F\otimes \mathcal E_0)$ is measurable, where $\chi^{-1}(\mathcal E_0)=\{\chi^{-1}(B)\,|\,B\in\mathcal E_0\}$ is a $\sigma$-algebra on $\Omega$;
\item 
if the map $Y(\cdot): \Omega\times E\rightarrow \mathbb R$ is $\mathcal F\otimes \mathcal E_0$-measurable, then $Y(\chi):\,\Omega\rightarrow \mathbb R$ is $\mathcal F\vee\chi^{-1}(\mathcal E_0)$-measurable.
\end{enumerate} 
\end{Lem}
\proof For 1), it suffices to prove that for all $A\in\mathcal F$ and $B\in\mathcal E_0$, one has  $$\Gamma_{\chi}^{-1}(A\times B)\in \mathcal F\vee\chi^{-1}(\mathcal E_0)$$ since $\mathcal F\otimes\mathcal E_0$ is generated by the sets of the form $A\times B$. Indeed,
\[\begin{split}\Gamma_{\chi}^{-1}(A\times B)&=
\{\omega\in\Omega \,|\, (\omega,\chi(\omega))\in A\times B\}=\{\omega\in A\,|\,\chi(\omega)\in B\}\\&=A\cap\chi^{-1}(B)\in\mathcal F\vee\chi^{-1}(\mathcal E_0)\end{split}\]which implies the first assertion. The second assertion 2) results from the fact that the composition of two measurable maps is still measurable.
\finproof
The above Lemma \ref{Lem:measurability} implies directly the following result.
\begin{Cor}Let $(Y_t(\cdot), t\geq 0)$ be a process adapted to the filtration $\mathbb F\otimes\mathcal N^E$, then $(Y_t(\chi),t\geq 0)$ is a $\mathbb G$-adapted process.
\end{Cor}

The proposition below calculates the $\mathbb G$-conditional expectations, which generalizes \cite[Theorem 3.1]{ejj1} for classic progressive enlargement of filtration and  \cite[Proposition 2.2]{ejj2} for successive multiple default times. It provides a very concise formula for computations and applications (the formula has the same form as  in the particular  case of product space in Section \ref{subsection: product space}). Moreover, we show that to make estimations with respect to the filtration $\mathbb G$, the key terms are the prediction process and the Radon-Nikodym derivative.

\begin{Pro}\label{Pro:espcond}
Let $Y_T(\cdot)$ be a non-negative $\mathcal F_T\otimes\mathcal E$-measurable function on $\Omega\times E$ and $t\leq T$. Then
\begin{equation}\label{equ:espcondg with density general}\mathbb E_{\mathbb P}[Y_T(\chi)|\mathcal G_t]=\frac{\int_E\mathbb E_{\mathbb P}[Y_T(x)\beta_T(x)|\mathcal F_t]\,\eta_t(dx)}{\int_E\beta_t(x)\,\eta_t(dx)},\end{equation}
where $\eta_t$ is the conditional law of $\chi$ given $\mathcal N_t$ and $\beta_t(x)$ is as in \eqref{equ:beta def}. 
\end{Pro}
\proof
Recall that $\overline{\proba}$ denotes the product probability measure $\mathbb P\otimes\eta$ on $(\Omega\times E,\mathcal A\otimes\mathcal E)$. Then for any $t\geq 0$,  the probability $\mathbb P'$  defined in \eqref{equ:P'} is absolutely continuous with respect to $\overline{\proba}$ on $\mathcal F_t\otimes\mathcal E$ with the Radon-Nikodym derivative given by $\beta_t(\cdot)$.
Indeed, if $f$ is a non-negative $\mathcal F_t\otimes\mathcal E$-measurable function, then by definition \eqref{equ:P'}, the expectation of $f$ with respect to the probability measure $\mathbb P'$ is 
\begin{equation}\label{radon nikodym P' and P bar}\begin{split}&\quad\int_{\Omega\times E} f(\omega,x)\mathbb P'(d\omega,dx)=\mathbb E_{\mathbb P}[f(\cdot,\chi)]=\mathbb E_{\mathbb P}[\mathbb E_{\mathbb P}[f(\cdot,\chi)|\mathcal F_t]]\\
&=\mathbb E_{\mathbb P}\bigg[\int_Ef(\cdot,x)\beta_t(x)\,\eta(dx)\bigg]=\int_{\Omega\times E}f(\omega,x)\beta_t(x)\,\overline{\proba}(d\omega,dx).\end{split}\end{equation}

We next consider a non-negative $\mathcal A\otimes\mathcal E$-measurable random variable $Y(\cdot)$ on $\Omega\times E$. By Lemma \ref{Lem:measurability} and the definition of $\mathbb P'$, for any sub-$\sigma$-algebra $\mathcal F$ of $\mathcal A$ and any sub-$\sigma$-algebra $\mathcal E_0$ of $\mathcal E$, we have 
\begin{equation}\label{equ:lemma 4.6}\mathbb E_{\mathbb P}[Y(\chi)|\mathcal F\vee\chi^{-1}(\mathcal E_0)]=\mathbb E_{\mathbb P'}[Y(\cdot)|\mathcal F\otimes\mathcal E_0](\chi)\end{equation} 
where the expression $\mathbb E_{\mathbb P'}[Y(\cdot)|\mathcal F\otimes\mathcal E_0](\chi)$ denotes the composition $\mathbb E_{\mathbb P'}[Y(\cdot)|\mathcal F\otimes\mathcal E_0]\circ\Gamma_{\chi}$ as indicated by \eqref{equ:composition}. Hence, we obtain for $\mathcal G_t=\mathcal F_t\vee\chi^{-1}(\mathcal N_t^E)$ the equality
\begin{equation}\label{equ: g_t esperance cond par proba produit}\mathbb E_{\mathbb P}[Y(\chi)|\mathcal G_t]=\mathbb E_{\mathbb P'}[Y(\cdot)|\mathcal F\otimes\mathcal N_t^E](\chi).\end{equation}

Finally, we obtain  by \eqref{equ: g_t esperance cond par proba produit} and \eqref{radon nikodym P' and P bar} that
\[\mathbb E_{\mathbb P}[Y_T(\chi)|\mathcal G_t]=\mathbb E_{\mathbb P'}[Y_T(\cdot)|\mathcal F_t\otimes\mathcal N_t^E](\chi)=\frac{\mathbb E_{\overline{\mathbb P}}\big[Y_T(\cdot)\beta_T(\cdot)|\mathcal F_t\otimes\mathcal N_t^E\big]}{\mathbb E_{\overline{\mathbb P}}\big[\beta_t(\cdot)|\mathcal F_t\otimes\mathcal N_t^E\big]}(\chi),\quad\mathbb P\text{-a.s.}\]
which implies the equality \eqref{equ:espcondg with density general} since $\overline{\proba}$ is the product probability measure $\mathbb P\otimes\eta$. 

\finproof

\begin{Rem}Similarly as noted in Remark \ref{beta positive},  on the set $\{\int_E\beta_t(x)\,\eta_t(dx)=0\}$, one has $\int_E\mathbb E_{\mathbb P}[Y_T(x)\beta_T(x)|\mathcal F_t]\,\eta_t(dx)=0$, $\mathbb P$-a.s., and we omit the indicator $\indic_{\{\int_E\beta_t(x)\,\eta_t(dx)>0\}}$ on the right-hand side of \eqref{equ:espcondg with density general}. The same rule will also be applied to what follows.
\end{Rem}
In the following, we apply the above proposition to several particular cases which were presented in Example \ref{example 1}.

{\bf Case of one default:} We consider the case  where $\chi=\tau$ and the filtration $\mathbb G$ is the standard progressive enlargement of $\mathbb F$ by $\tau$. 
Then the $\mathcal N_t$-conditional law of $\tau$ is given by
\[\eta_t(du)=\frac{\indic_{\{t<u\}}\alpha_0(u)\nu(du)}{\int_t^\infty\alpha_0(u)\nu(du)}\indic_{\{\tau>t\}}+{\delta_\tau(du)}\indic_{\{\tau\leq t\}}\] 
For any  non-negative $\mathcal F_T\otimes\mathcal B(\mathbb R_+)$-measurable function $Y_T(\cdot)$, one has (c.f. \cite[Theorem 3.1]{ejj1})
\begin{equation}\label{exp one default}\mathbb E_{\mathbb P}[Y_T(\tau)|\mathcal G_t]=\frac{\int_t^{\infty}\mathbb E_{\mathbb P}[Y_T(u)\alpha_T(u)|\mathcal F_t]\nu(du)}{\int_t^{\infty}\alpha_t(u)\nu(du)}\indic_{\{\tau>t\}}+\frac{\mathbb E_{\mathbb P}[Y_T(u)\alpha_T(u)|\mathcal F_t]}{\alpha_t(u)}\Big|_{u=\tau}\indic_{\{\tau\leq t\}}.\end{equation}

{\bf One default with insider information} Consider one default with insider's information  as in Example \ref{example 1} (3) where the filtration $\mathbb G$ is given by $(\mathcal F_t\vee\mathcal N_t)_{t\geq 0}$ with $\mathcal N_t=\sigma(\indic_{\{\tau\leq s\}},\, s\leq t\text{ or }s=t_0)$. Then the $\mathcal N_t$-conditional law of $\tau$ is given, according to $t<t_0$ or $t\geq t_0$, by
\[\eta_t(du)=\frac{\indic_{\{t<u\leq t_0\}}\alpha_0(u)\nu(du)}{\int_t^{t_0}\alpha_0(u)\nu(du)}\indic_{\{t<\tau\leq t_0\}}+\frac{\indic_{\{t_0<u\}}\alpha_0(u)\nu(du)}{\int_{t_0}^\infty\alpha_0(u)\nu(du)}\indic_{\{\tau>t_0\}}+\delta_{\tau}(du)\indic_{\{\tau\leq t\}},\text{ if $t<t_0$}\]
\[\eta_t(du)=\frac{\indic_{\{t<u\}}\alpha_0(u)\nu(du)}{\int_t^\infty\alpha_0(u)\nu(du)}\indic_{\{\tau>t\}}+{\delta_\tau(du)}\indic_{\{\tau\leq t\}},\text{ if $t\geq t_0$}.\] 
For any  non-negative $\mathcal F_T\otimes\mathcal B(\mathbb R_+)$-measurable function $Y_T(\cdot)$, one has respectively
\[\begin{split}\mathbb E_{\mathbb P}[Y_T(\tau)|\mathcal G_t]&=\frac{\int_t^{t_0}\mathbb E_{\mathbb P}[Y_T(u)\alpha_T(u)|\mathcal F_t]\nu(du)}{\int_t^{t_0}\alpha_t(u)\nu(du)}\indic_{\{t_0\geq \tau>t\}}\\
&+\frac{\int_{t_0}^{\infty}\mathbb E_{\mathbb P}[Y_T(u)\alpha_T(u)|\mathcal F_t]\nu(du)}{\int_{t_0}^{\infty}\alpha_t(u)\nu(du)}\indic_{\{\tau>t_0\}}+\frac{\mathbb E_{\mathbb P}[Y_T(u)\alpha_T(u)|\mathcal F_t]}{\alpha_t(u)}\Big|_{u=\tau}\indic_{\{\tau\leq t\}},\text{ if $t< t_0$}.
\end{split}\]
\[\mathbb E_{\mathbb P}[Y_T(\tau)|\mathcal G_t]=\frac{\int_t^{\infty}\mathbb E_{\mathbb P}[Y_T(u)\alpha_T(u)|\mathcal F_t]\nu(du)}{\int_t^{\infty}\alpha_t(u)\nu(du)}\indic_{\{\tau>t\}}+\frac{\mathbb E_{\mathbb P}[Y_T(u)\alpha_T(u)|\mathcal F_t]}{\alpha_t(u)}\Big|_{u=\tau}\indic_{\{\tau\leq t\}},\text{ if $t\geq t_0$}.\]

We note that when $t\geq t_0$, the information flow becomes the same as for a standard investor, so that the conditional expectation formula in this case coincides with \eqref{exp one default}.


{\bf Multiple ordered defaults:} $\chi=\bsigma=(\sigma_1,\cdots, \sigma_n)$ where $\sigma_1\leq \cdots\leq\sigma_n$ and $E=\{(u_1,\cdots,u_n)\in\mathbb R_+^n\,|\,u_1\leq\cdots\leq u_n\}$. The filtration $(\mathcal N_t)_{t\geq 0}$ is generated by the process $(\sum_{i=1}^n\indic_{\{\sigma_i\leq t\}},\;t\geq 0)$. Assume  that $\nu$ is the Lebesgue measure and that the $\mathcal F_t$-conditional law of $\chi$ has a density $\alpha_t(\cdot)$ with respect to $\nu(d\boldsymbol{u})=d\boldsymbol{u}$. Then the $\mathcal N_t$-conditional law is given by 
\begin{equation}\label{equ:eta N ordered defaults}\eta_t(d\bu)=\sum_{i=0}^n\frac{\indic_{\{t<u_{i+1}\}}\alpha_0(\bsigma_{(i)},\bu_{(i+1:n)})\delta_{(\bsigma_{(i)})}(d\bu_{(i)})d{\bu_{(i+1:n)}}}{\int_t^{\infty}\alpha_0(\bsigma_{(i)},\bu_{(i+1:n)})d\bu_{(i+1:n)}}\,\indic_{E^i_t}(\bsigma)\end{equation}
where 
\begin{equation}\label{equ E i}E^i_t:=\{(u_1,\ldots,u_n)\in E\,|\,u_i\leq t<u_{i+1}\}.\end{equation}
Then by Proposition \ref{Pro:espcond}, we obtain  $$\esp_{\proba}[Y_T(\bsigma)|\G_t]=\sum_{i=0}^{n}\indic_{\{\sigma_i\leq t<\sigma_{i+1}\}} \frac{\int_t^\infty\esp_{\proba}[Y_T(\bu)\alpha_T(\bu)|\F_t]\,d\bu_{(i+1: n)}}{\int_t^{\infty}\alpha_{t}(\bu)\,d\bu_{(i+1: n)}}\Big|_{\bu_{(i)}=\bsigma_{(i)}}$$
which corresponds to  \cite[Proposition 2.2]{ejj2}.

{\bf Multiple non-ordered defaults:} $\chi=\btau=(\tau_1,\cdots, \tau_n)$ and $E=\mathbb R_+^n$. The filtration $(\mathcal N_t)_{t\geq 0}$ is generated by the family of indicator processes $(\indic_{\{\tau_i\leq t\}},\;t\geq 0)$, $i=1,\cdots, n$. Assume in addition that the $\mathcal F_t$-conditional law of $\chi$ has a density $\alpha_t(\cdot)$ with respect to the Lebesgue measure $d\boldsymbol{u}$. Then the $\mathcal N_t$-conditional law of $\eta$ can be written in the form  
\begin{equation}\label{equ:eta N non-ordered defaults}\eta_t(d\bu)=\sum_{I\subset\{1,\cdots, n\}}\frac{\indic_{\{\bu_{I^c}>t\}}\alpha_0(\cdot,\bu_{I^c}) \delta_{(\cdot)}(d\bu_I)d\bu_{I^c}}{\int_t^{\infty}\alpha_0(\cdot,\bu_{I^c})d\bu_{I^c}}\,\indic_{E_t^I}(\btau),\end{equation}where for $I\subset\{1,\ldots,n\}$, $\delta_{(\cdot)}(d\bu_I)$ denotes the Dirac measure on the coordinates with indices in $I$, $\bu_{I^c}$ denotes the vector $(u_j)_{j\in I^c}$,  the event $\{\bu_{I^c}>t\}$ denotes $\bigcap_{j\in I^c}\{\bu_j>t\}$, and 
\[E_t^I:=\{(u_1,\ldots,u_n)\in E\,|\,\forall\,i\in I,\,u_i\in[0,t],\;\forall\,j\in I^c,\,u_j>t\}.\] 
By Proposition \ref{Pro:espcond}, we obtain
\[\mathbb E_{\mathbb P}[Y_T(\btau)|\mathcal G_t]=\sum_{I\in\{1,\cdots, n\}}\indic_{\{\btau_I\leq t,\,\btau_{I^c}>t\}}\frac{\int_t^{\infty}\mathbb E_{\mathbb P}[Y_T(\bu)\alpha_T(\bu)|\mathcal F_t]d\bu_{I^c}}{
\int_t^{\infty}\alpha_t(\bu)d\bu_{I^c}}\Big|_{\bu_I=\btau_I}\]
where $\btau_I:=(\tau_i)_{i\in I}$ and $\indic_{\{\btau_I\leq t,\,\btau_{I^c}>t\}}$ corresponds to $\indic_{E_t^I}(\btau)$. 

If the ordered defaults $\bsigma$ is defined as the increasing permutation of $\btau$, then there exists an explicit relation between the density processes $\alpha^{\bsigma}(\cdot)$ of $\bsigma$ and $\alpha^{\btau}(\cdot)$ of $\btau$ by using the order statistics. For any $\bu\in\mathbb R_+^n$ such that $u_1<\cdots<u_n$, one has for any $t\geq 0$, 
\[\alpha^{\bsigma}_t(u_1,\cdots,u_n)=\indic_{\{u_1<\cdots<u_n\}}\sum_{\Pi}\alpha^{\btau}_t(u_{\Pi(1)},\cdots, u_{\Pi(n)})\]
where $(\Pi(1),\cdots,\Pi(n)$ is a permutation of $(1,\cdots, n)$.  If in addition $\btau$ is exchangeable (see e.g. \cite{FM2001}), then for any permutation, $(\tau_{\Pi(1)},\cdots,\tau_{\Pi(n)})$ has the same distribution as $(\tau_1,\cdots,\tau_n)$ so that \[\alpha^{\bsigma}_t(u_1,\cdots,u_n)=\indic_{\{u_1<\cdots<u_n\}}n!\,\alpha^{\btau}_t(u_1,\cdots, u_n).\] In this case we say that the default portfolio is homogeneous. 

\section{Martingale characterization}\label{subsection: martingale}
It is important to study martingale properties for financial applications such as pricing of credit sensitive contingent claims. In this section, we are interested in the  characterization of  martingale processes in different enlarged  filtrations, notably in the observation information filtration $\mathbb G$.  

We first recall a martingale criterion in the initial enlargement of filtration  in Amendinger \cite{Am2000} (see also Callegaro, Jeanblanc and Zargari \cite{CJZ}). It corresponds in our setting to the total information filtration $\mathbb H=(\mathcal H_t)_{t\geq 0}$, $\mathcal H_t=\cap_{s>t}(\mathcal F_s\vee\sigma(\chi))$. For the ease of readers, we also give the proof below. 

\begin{Pro} \label{pro: martingale initiale} An $(\mathcal F_t\otimes\mathcal E)_{t\geq 0}$-adapted process $(M_t(\cdot),t\geq 0)$ is an $(\mathcal F_t\otimes\mathcal E)_{t\geq 0}$-martingale under the probability measure $\mathbb P'$ defined in \eqref{equ:P'} if and only if $(\alpha_t(x)M_t(x), t\geq 0)$ is a parametrized $({\mathbb F},{\mathbb P})$-martingale depending on ${x\in E}$. Moreover, if this condition is satisfied, then $(M_t(\chi),t\geq 0)$ is an $(\mathbb H,\mathbb P)$-martingale on $(\Omega,\mathcal A)$.
\end{Pro}
\proof
For any $\mathcal F_T\otimes\mathcal E$-measurable random variable $M_T(\cdot)$ and $t\leq T$, since the Radon-Nikodym derivative of $\mathbb P'$ with respect to $\overline{\mathbb P}$ is $\beta_T(\cdot)$ on $\mathcal F_T\otimes\mathcal E$, we have that
\[{\alpha_t(\cdot)}\mathbb E_{\mathbb P'}[M_T(\cdot)|\mathcal F_t\otimes\mathcal E]=\mathbb E_{\overline{\mathbb P}}\big[{M_T(\cdot)\alpha_T(\cdot)}\,\big|\,\mathcal F_t\otimes\mathcal E\big]={\mathbb E_{\mathbb P}[M_T(\cdot)\alpha_T(\cdot)|\mathcal F_t]}.\]
Note that $\alpha_t(\cdot)>0$ almost surely under $\mathbb P'$. In fact, by \eqref{equ:P'} one has
\[\mathbb E_{\mathbb P'}[\indic_{\{\alpha_t(\cdot)=0\}}]=\mathbb E_{\mathbb P}[\indic_{\{\alpha_t(\chi)=0\}}]=\mathbb E_{\mathbb P}\bigg[\int_E\indic_{\{\alpha_t(x)=0\}}\alpha_t(x)\nu(dx)\bigg]=0.\]
Therefore, the process $M(\cdot)$ is an $(\mathbb P',\mathcal F_t\otimes\mathcal E)_{t\geq 0}$-martingale if and only if $(\alpha_t(x)M_t(x), t\geq 0)_{x\in E}$ is a parametrized $({\mathbb F},{\mathbb P})$-martingale depending on  $x\in E$. Finally, \eqref{equ:lemma 4.6} implies that
\[\mathbb E_{\mathbb P}[M_T(\chi)|\mathcal H_t]=\mathbb E_{\mathbb P'}[M_T(\cdot)|\mathcal F_t\otimes\mathcal E](\chi).\]
Therefore we obtain the second assertion of the proposition.
\finproof

\subsection{Martingales in the filtration $\mathbb G$}
We now deduce from Proposition \ref{Pro:espcond} the following martingale criterion for the accessible information filtration $\mathbb G$. \begin{Thm}\label{thm: martingale G}
Let $(M_t(\cdot),t\geq 0)$ be $(\mathcal F_t\otimes\mathcal N_t^E)_{t\geq 0}$-adapted processes. If the process
\begin{equation*}\widetilde M_t(\chi)=M_t(\chi)\int_E\beta_t(x)\eta_t(dx),\quad t\geq 0\end{equation*} verifies
\begin{equation}\label{Equ:Mtild}\forall\, T\geq t\geq 0,\quad \int_{E}\mathbb E_{\mathbb P}[\widetilde M_T(x)|\mathcal F_t]\,\eta_t(dx)=\widetilde M_t(\chi),\end{equation}
then $(M_t(\chi),t\geq 0)$ is a $(\mathbb G,\mathbb P)$-martingale.

\end{Thm}
\proof For any $t\geq 0$, let $\eta_t^E$ be the  conditional law of   $\chi$ given $\mathcal N_t^E$ on $(E,\mathcal E)$, i.e., for any bounded or non-negative Borel function $f$ on $E$, \begin{equation}\label{eta^E}\int_Ef(x)\eta_t(dx)=\bigg(\int_E f(x)\eta_t^E(dx)\bigg)(\chi).\end{equation} 
Then, by Proposition \ref{Pro:espcond}, for $T\geq t\geq 0$,
\[\mathbb E_{\mathbb P}[M_T(\chi)|\mathcal G_t]=\frac{\int_E\mathbb E_{\mathbb P}[M_T(x)\beta_T(x)|\mathcal F_t]\eta_t^E(dx)}{\int_E\beta_t(x)\eta_t^E(dx)}(\chi).\]
By Fubini's theorem for conditional expectations, 
\begin{equation}\label{equ:numerateur1}\int_E\mathbb E_{\mathbb P}\Big[M_T(x)\beta_T(x)\,\Big|\,\mathcal F_t\Big]\eta_t^E(dx)=\mathbb E_{\overline{\mathbb P}}\Big[M_T(\cdot)\beta_T(\cdot)\,\Big|\,\mathcal F_t\otimes\mathcal N_t^E\Big]\end{equation}
Note that by \eqref{radon nikodym P' and P bar}, $d\mathbb P'=\beta_T(\cdot)d\overline{\mathbb P}$ on $\mathcal F_T\otimes\mathcal E$. Since $\mathbb P'$ is induced by the graph map $\Gamma_\chi$, for any bounded or non-negative $\mathcal F_T\otimes\mathcal E$-measurable random variable $\varphi_T(\cdot)$, one has \[\mathbb E_{\overline{\mathbb P}}\Big[\varphi_T(\cdot)\beta_T(\cdot)\Big]=\mathbb E_{\mathbb P'}[\varphi_T(\cdot)]=\mathbb E_{\mathbb P}[\varphi_T(\chi)],\] which only depends on the value of the random variable $\varphi_T(\chi)$ on $\Omega$. Hence the random variable $\mathbb E_{\overline{\mathbb P}}[M_T(\cdot)\beta_T(\cdot)\,|\,\mathcal F_t\otimes\mathcal N_t^E]$, which is the right-hand side of the equality \eqref{equ:numerateur1}, only depends on the value of $M_T(\chi)$. Therefore in the computation of \eqref{equ:numerateur1}, we may assume without loss of generality that
\begin{equation}\label{Equ:widetilde Mtchi}\widetilde M_T(\cdot)=M_T(\cdot)\int_E\beta_T(x)\eta_T^E(dx).\end{equation}
By Bayes' formula, one has
\begin{equation}\label{equ:changetop'}\int_E\mathbb E_{\mathbb P}\Big[M_T(x)\beta_T(x)\,\Big|\,\mathcal F_t\Big]\eta_t^E(dx)=\mathbb E_{\overline{\mathbb P}}\Big[M_T(\cdot)\int_E\beta_T(x)\eta^E_T(dx)\,\Big|\,\mathcal F_t\otimes\mathcal N_t^E\Big],\end{equation}
which by \eqref{Equ:widetilde Mtchi} leads to 
\[\int_E\mathbb E_{\mathbb P}\Big[M_T(x)\beta_T(x)\,\Big|\,\mathcal F_t\Big]\eta_t^E(dx)=\int_E\mathbb E_{\mathbb P}[\widetilde M_T(x)|\mathcal F_t]\eta_t^E(dx).\]
Finally the condition \eqref{Equ:Mtild} implies 
\[\mathbb E_{\mathbb P}[M_T(\chi)|\mathcal G_t]=\frac{\widetilde M_t(\chi)}{\big(\int_E\beta_t(x)\eta_t(dx)\big)(\chi)}=M_t(\chi).\]
The theorem is thus proved.
\finproof

\begin{Rem}
The condition \eqref{Equ:Mtild} is satisfied notably when 
\[\int_E\mathbb E_{\mathbb P}[\widetilde M_T(x)|\mathcal F_t]\eta_t^E(dx)=\widetilde M_t(x)\]
for any $x\in E$, which means that $(\widetilde M_t(\cdot))_{t\geq 0}$ is an $(\mathcal F_t\otimes\mathcal N_t^E)_{t\geq 0}$-martingale under the product measure $\overline{\mathbb P}$. This observation allows to construct $(\mathbb G,\mathbb P)$-martingales. We begin with a $((\mathcal F_t\otimes\mathcal N_t^E)_{t\geq 0},\overline{\mathbb P})$-martingale $\widetilde M(\cdot)$ (which could be chosen easily as conditional expectations  on the product probability space $\Omega\times E$ since $\overline{\mathbb P}$ is the product measure). Then the process
\[M_t(\chi):=\widetilde M_t(\chi)\Big(\int_E\beta_t(x)\eta_t(dx)\Big)^{-1},\quad t\geq 0\]
is a $(\mathbb G,\mathbb P)$-martingale.\end{Rem}

\begin{Cor}Given a strictly positive $(\mathbb G,\mathbb P)$-martingale $(M_t(\chi),t\geq 0)$ with $\mathbb P$-expectation $1$, let $\mathbb Q$ be an equivalent probability measure of $\mathbb P$ defined by 
\[\frac{d\mathbb Q}{d\mathbb P}\,\big|_{\mathcal G_t}=M_t(\chi).\]
Then, for any non-negative Borel function $f(\cdot)$ on $E$, one has
\[\mathbb E_{\mathbb Q}[f(\chi)|\mathcal F_t]=\frac{\mathbb E_{\mathbb P}[f(\chi)M_t(\chi)|\mathcal F_t]}{\mathbb E_{\mathbb P}[M_t(\chi)|\mathcal F_t]}=\frac{\int_Ef(x)M_t(x)\alpha_t(x)\nu(dx)}{\int_EM_t(x)\alpha_t(x)\nu(dx)}.\]
In other words,  the $\mathcal F_t$-conditional density of $\chi$ under $\mathbb Q$ is
\[\alpha_t^{\mathbb Q}(x)=\frac{M_t(x)\alpha_t(x)}{\int_EM_t(x)\alpha_t(x)\nu(dx)}.\]
\end{Cor}

The immersion property between  filtrations $\mathbb F$ and $\mathbb G$ asserts that any $\mathbb F$-martingale remains a $\mathbb G$-martingale. We present below a direct consequence of Theorem \ref{thm: martingale G} concerning the immersion property.
\begin{Cor}Assume that, for $0\leq t\leq T$, one has
\begin{equation}\label{equ:Hhypothesis}\int_E\beta_t(x)\eta_t^E(dx)=\int_E\beta_T(x)\eta_t^E(dx),\end{equation}
then $(\mathbb F,\mathbb G)$ satisfies the immersion property.
\end{Cor}
\proof Let $M$ be an $(\mathbb F,\mathbb P)$-martingale. For $t\geq 0$ and $x\in E$, let
\[\widetilde M_t(x)=M_t\int_E\beta_t(x)\eta_t^E(dx).\]
One has
\[\mathbb E_{\mathbb P}[\widetilde M_T(x)|\mathcal F_t]\eta_t^E(dx)=\mathbb E_{\mathbb P}\bigg[M_T\int_E\beta_T(x)\eta_t^E(dx)\,\bigg|\,\mathcal F_t\bigg]=M_t\int_E\beta_t(x)\eta_t^E(dx),\]
where the last equality comes from \eqref{equ:Hhypothesis}. By Theorem \ref{thm: martingale G}, we obtain that $M$ is a $(\mathbb G,\mathbb P)$-martingale.
\finproof

\subsection{Special cases of ordered and non-ordered multi-defaults}
We apply the martingale characterization result to several particular cases. In the case of single default when $\chi=\tau$ and $E=\mathbb R_+$, Theorem \ref{thm: martingale G} leads to  \cite[Theorem 5.7]{ejj1}. In the following, we give some extensions for multi-default cases by using Theorem \ref{thm: martingale G}.

{\bf Case of ordered defaults:} This case can be viewed as a generalization of the single default case. 
By \eqref{equ:eta N ordered defaults}, we have for any $\boldsymbol{v}\in E=\{(v_1,\cdots,v_n)\in\mathbb R_+^n\,|\,v_1\leq\cdots\leq v_n\}$, 
\[\Big(\int_E{\beta_t(\bu)}\eta_t(d\bu)\Big)(\boldsymbol{v})=
\sum_{i=0}^n\frac{\int_t^{\infty}\alpha_t(\boldsymbol{v}_{(i)},\bu_{(i+1:n)})d\bu_{(i+1:n)}}{\int_t^{\infty}\alpha_0(\boldsymbol{v}_{(i)},\bu_{(i+1:n)})d\bu_{(i+1:n)}}\indic_{E_t^i}(\boldsymbol{v}).\]
Since $(\mathcal N_t^E)_{t\geq 0}$ is generated by the process $(N_t=\sum_{i=1}^n\indic_{\{u_i\leq t\}},\;t\geq 0)$, the $\mathbb F\otimes\mathcal N^E$-adapted process $M(\bu)$ can be  written in the form
\[M_t(\bu)=\sum_{i=0}^nM_t^i(\bu_{(i)})\indic_{E_t^i}(\bu),\quad t\geq 0\]
where $M^i(\cdot)$ is $\mathbb F\otimes\mathcal B(\mathbb R_+^i)$-adapted and $E_t^i$ is defined in \eqref{equ E i}, then one has
\[\widetilde M_t(\bu)=\sum_{i=0}^n\bigg(M_t^i(\bu_{(i)})\frac{\int_t^{\infty}\alpha_t(\bu)d\bu_{(i+1:n)}}{\int_t^\infty\alpha_0(\bu)d\bu_{(i+1:n)}}\bigg)\indic_{E_t^i}(\bu).\]
Therefore, for $T\geq t\geq 0$, 
\[\mathbb E_{\mathbb P}[\widetilde M_T(\bu)|\mathcal F_t]=\sum_{i=0}^{n}\frac{\mathbb E_{\mathbb P}[M_T^i(\bu_{(i)})\int_T^\infty\alpha_T(\bu)d\bu_{(i+1:n)}|\mathcal F_t]}{\int_{T}^\infty\alpha_0(\bu)d\bu_{(i+1:n)}}\indic_{E_T^i}(\bu)\]
and
\[\begin{split}&\quad\;\Big(\int_E\mathbb E_{\mathbb P}[\widetilde M_T(\bu)|\mathcal F_t]\eta_t(d\bu)\Big)(\boldsymbol{v})\\
&=\sum_{j=0}^n\Big(\sum_{i\geq j}\frac{\int_t^T\mathbb E_{\mathbb P}[M_T^i(\boldsymbol{v}_{(i)})\int_T^\infty\alpha_T(\boldsymbol{v}_{(i)},\bu_{(i+1:n)})d\bu_{(i+1:n)}|\mathcal F_t]d\boldsymbol{v}_{(j+1:i)}}{\int_t^\infty\alpha_0(\boldsymbol{v})d\boldsymbol{v}_{(j+1:n)}}\Big)\indic_{E_t^j}(\boldsymbol{v}).
\end{split}\]
So the condition \eqref{Equ:Mtild} is  equivalent to, for any $j\in\{0,\ldots,n\}$,
\begin{equation}\label{Equ:egalitej}
\sum_{i\geq j}\int_t^T\mathbb E_{\mathbb P}\Big[M_T^i(\boldsymbol{u}_{(i)})\int_T^\infty\alpha_T(\bu)d\bu_{(i+1:n)}\Big|\mathcal F_t\Big]d\boldsymbol{u}_{(j+1:i)}=M_t^{j}(\boldsymbol{u}_{(j)})\int_t^\infty\alpha_t(\boldsymbol{u})d\boldsymbol{u}_{(j+1:n)}, \,t\geq u_j\end{equation}
which implies the following characterization result.

\begin{Pro}\label{pro:martingal charc ordered}
The condition \eqref{Equ:Mtild} is equivalent to the following: for any $j\in\{0,\ldots,n\}$ and any $\boldsymbol{u}_{(j)}\in\mathbb R^j_+$, $u_1\leq\cdots\leq u_j$,
\begin{equation}\label{Equ:martigalej}M_t^j(\boldsymbol{u}_{(j)})\int_t^\infty\alpha_t(\boldsymbol{u})d\bu_{(j+1:n)}-\int_0^tM_{u_{j+1}}^{j+1}(\bu_{(j+1)})\int_{u_{j+1}}^{\infty}\alpha_{u_{j+1}}(\bu)d\bu_{(j+2:n)}du_{j+1},\,\,\, t\geq u_j\end{equation}
is an $(\mathbb F,\mathbb P)$-martingale.
\end{Pro}
\proof For any $j\in\{0,\cdots, n\}$, let $(A_j)$ be the equality \eqref{Equ:egalitej} for $T\geq t\geq 0$ and $\bu_{(j)}=(u_1,\cdots,u_j)\in\mathbb R^j_+$ such that $u_1\leq\cdots\leq u_j\leq t$ and let $(B_j)$ be the martingale property of \eqref{Equ:martigalej}. We will prove by reverse induction on $j$ that
\begin{equation}\label{Equ:equivalencej}(\forall\,i\geq j,\;(A_i))\Longleftrightarrow (\forall\,i\geq j,\;(B_i)).\end{equation} Note that the conditions ($A_n$) and $(B_n)$ are acutally the same. Assume that the equivalence \eqref{Equ:equivalencej} has been proved for $j'\geq j$. We will prove the equivalence for $j$. By the induction assumption it suffice to prove $(A_j)\Leftrightarrow(B_j)$ given  that $(A_i)$ and $(B_i)$ are satisfied for all $i>j$.
Thus for $i\geq j+1$ and $t<u_{j+1}$ one has
\[\begin{split}&\quad\;\mathbb E_{\mathbb P}\Big[M_T^i(\boldsymbol{u}_{(i)})\int_T^\infty\alpha_T(\boldsymbol{u})d\bu_{(i+1:n)}\Big|\mathcal F_t\Big]=\mathbb E_{\mathbb P}\Big[\mathbb E\Big[M_T^i(\boldsymbol{u}_{(i)})\int_T^\infty\alpha_T(\boldsymbol{u})d\bu_{(i+1:n)}\Big|\mathcal F_{u_i}\Big]\mathcal F_t\Big]\\&=\mathbb E_{\mathbb P}\Big[M_{u_i}^i(\boldsymbol{u}_{(i)})\int_{u_i}^\infty\alpha_{u_i}(\bu)d\bu_{(i+1:n)}\Big|\mathcal F_t\Big]-\int_{u_{i}}^T\mathbb E_{\mathbb P}\Big[M_{u_{i+1}}^{i+1}(\bu_{(i+1)})\int_{u_{i+1}}^\infty\alpha_{u_{i+1}}(\bu)d\bu_{(i+2:n)}\Big|\mathcal F_t\Big]du_{i+1}.\end{split}\]
Therefore the equality \eqref{Equ:egalitej} is equivalent to 
\[\begin{split}&\quad\;\mathbb E_{\mathbb P}\Big[M_T^j(\bu_{(j)})\int_T^\infty\alpha_T(\bu)d\bu_{(i+1:n)}\Big|\mathcal F_t\Big]-M_t^j(\bu_{(j)})\int_t^\infty\alpha_t(\bu)d\bu_{(j+1:n)}\\
&=\sum_{i\geqslant j+1}\bigg(\int_t^T\mathbb E_{\mathbb P}\Big[M_{u_i}^{i}(\bu_{(i)})\int_{u_i}^\infty\alpha_{u_i}(\bu)d\bu_{(i+1:n)}\Big|\mathcal F_t\Big]d\bu_{(j+1:i)}\\
&\qquad-\int_t^T\mathbb E_{\mathbb P}\Big[M_{u_{i+1}}^{i+1}(\bu_{(i+1)})\int_{u_{i+1}}^\infty\alpha_{u_{i+1}}(\bu)d\bu_{(i+2:n)}\Big|\mathcal F_t\Big]d\bu_{(i+2:n)}\bigg)\\
&=\int_t^T\mathbb E_{\mathbb P}\Big[M_{u_{j+1}}^{j+1}(\bu_{(j+1)})\int_{u_{j+1}}^\infty\alpha_{u_{j+1}}(\bu)d\bu_{(j+2:n)}\Big|\mathcal F_t\Big]du_{j+1}.\end{split}\]
Hence we obtain the equivalence of $(A_j)$ and $(B_j)$.
\finproof

{\bf Case of non-ordered defaults:} The case of non-ordered defaults and ordered ones can be treated in similar way. 
The only difference is to make precise the corresponding prediction process. For $\boldsymbol{v}\in\mathbb R_+^n$, one has by \eqref{equ:eta N non-ordered defaults} that
\[\Big(\int_E{\beta_t(\bu)}\eta_t(d\bu)\Big)(\boldsymbol{v})=\sum_{I\subset\{1,\cdots,n\}}\frac{\int_t^\infty\alpha_t(\boldsymbol{v}_I,\bu_{I^c})d\bu_{I^c}}{\int_t^\infty\alpha_0(\boldsymbol{v}_I,\bu_{I^c})d\bu_{I^c}}\indic_{E_t^I}(\boldsymbol{v}).\]
The $\mathbb F\otimes\mathcal N^E$-adapted process $M(\bu)$ can be written in the form
\[M_t(\bu)=\sum_{I\subset\{1,\ldots,n\}}M_t^I(\bu_I)\indic_{E_t^I}(\bu),\quad t\geq 0\]
where $M^I(\cdot)$ is $\mathbb F\otimes\mathcal B(\mathbb R_+^I)$-adapted, then one has
\[\widetilde M_t(\bu)=\sum_{I\subset\{1,\ldots,n\}}\Big(M_t^I(\bu_I)\frac{\int_t^\infty\alpha_t(\bu)d\bu_{I^c}}{\int_t^\infty\alpha_0(\bu)d\bu_{I^c}}\Big)\indic_{E_t^I}(\bu)\]
and for $T\geq t\geq 0$, 
\[\mathbb E_{\mathbb P}[\widetilde M_T(\bu)|\mathcal F_t]=\sum_{I\subset\{1,\ldots,n\}}\frac{\mathbb E_{\mathbb P}[M_T^I(\bu_I)\int_T^\infty\alpha_T(\bu)d\bu_{I^c}|\mathcal F_t]}{\int_T^\infty\alpha_0(\bu)d\bu_{I^c}}\indic_{E_t^I}(\bu).\]
Therefore 
\[\begin{split}&\quad\;\Big(\int_E\mathbb E_{\mathbb P}[\widetilde M_T(\bu)|\mathcal F_t]\eta_t(d\bu)\Big)(\boldsymbol{v})\\
&=\sum_{J\subset\{1,\ldots n\}}\Big(\sum_{I\supset J}\int_t^{\infty}\frac{\mathbb E_{\mathbb P}[M_T^I(\boldsymbol{v}_I)\int_T^\infty\alpha_T(\boldsymbol{v}_I,\bu_{I^c})d\bu_{I^c}|\mathcal F_t]}{\int_T^\infty\alpha_0(\boldsymbol{v}_I,\bu_{I^c})d\bu_{I^c}} \indic_{E_T^I}(\boldsymbol{v})\alpha_0(\boldsymbol{v})d\boldsymbol{v}_{J^c}\Big)\frac{\indic_{E_t^J}(\boldsymbol{v})}{\int_t^\infty\alpha_0(\boldsymbol{v})d\boldsymbol{v}_{J^c}}\\
&=\sum_{J\subset\{1,\ldots,n\}}\Big(\sum_{I\supset J}\frac{\int_t^T\mathbb E_{\mathbb P}[M_T^I(\boldsymbol{v}_I)\int_T^\infty\alpha_T(\boldsymbol{v}_I,\bu_{I^c})d\bu_{I^c}|\mathcal F_t]d\boldsymbol{v}_{I\setminus J}}{\int_t^\infty\alpha_0(\boldsymbol{v})d\boldsymbol{v}_{J^c}}\Big)\indic_{E_t^J}(\boldsymbol{v})\end{split}\]
Therefore the conditon \eqref{Equ:Mtild} is actually equivalent to, for any $J\subset\{1,\ldots,n\}$, and $\bu_J\in \mathbb R_+^J$ such that $\bu_J^{\max}:=\displaystyle\max_{j\in J}u_j\leq t$,
\begin{equation}\label{Equ:martingalenonord}\sum_{I\supset J}\int_t^T\mathbb E_{\mathbb P}\Big[M_T^I(\boldsymbol{u}_I)\int_T^\infty\alpha_T(\bu)d\bu_{I^c}\Big|\mathcal F_t\Big]d\boldsymbol{u}_{I\setminus J}=M_t^J(\bu_J)\int_t^\infty\alpha_t(\bu)d\bu_{J^c}.\end{equation}

Similar as in the ordered case, we have the following characterization result for non-ordered defaults. The proof is analogous to that of Proposition \ref{pro:martingal charc ordered}. We therefore omit it.
\begin{Pro}
The condition \eqref{Equ:Mtild} is equivalent to the following: for any $J\subset\{0,\cdots,n\}$ and any $\bu_J\in\mathbb R_+^J$, the process
\begin{equation}\label{Equ:martingale}M_t^J(\bu_J)\int_t^\infty\alpha_t(\bu)d\bu_{J^c}-\sum_{k\in J^c}\int_{\bu_J^{\max}}^tM_{u_k}^{J\cup\{k\}}(\bu_{J\cup\{k\}})\Big(\int_{u_k}^\infty\alpha_{u_k}(\bu)d\bu_{J^c\setminus\{k\}}\Big)du_k,\;\,\,\bu_J^{\max}\leq t\end{equation}
is an $(\mathbb F,\mathbb P)$-martingale.
\end{Pro}

We finally give an example to show how to construct a $\mathbb G$-martingale when $n=2$ and $\chi=(\tau_1,\tau_2)$ by using the above proposition. 
\begin{Exe}
We define $M(\cdot,\cdot)$ to be an $(\mathcal F_t\otimes\mathcal N_t^E)_{t\geq 0}$-adapted process which is written in the form
\begin{equation}\label{equ:M}M_t(u_1,u_2)=\indic_{\{u_1>t,u_2>t\}}M^0_t+\sum_{\{i,j\}=\{1,2\}}\indic_{\{u_i\leq t,u_j> t\}}M^i_t(u_i)+\indic_{\{u_1\leq t,u_2\leq t\}}M_t^{1,2}(u_1,u_2),\end{equation}
where $M^0$ is $\mathbb F$-adapted, $M^1(\cdot)$ and $M^2(\cdot)$ are $\mathbb F\otimes\mathcal B(\mathbb R_+)$-adapted, and $M^{1,2}$ is $\mathbb F\otimes\mathcal B(\mathbb R_+^2)$-adapted processes.
Then $(M_t(\tau_1,\tau_2), t\geq 0)$ is a $\mathbb G$-adapted process. Let $(\alpha_t(u_1,u_2),t\geq 0)$ be the $\mathbb F$-conditional density process of the couple of non-ordered default times $(\tau_1,\tau_2)$, as defined in Hypothesis \ref{hyp:density}.
Let $(L^{1,2}_t(u_1,u_2),t\geq\max(u_1,u_2))$ be a family of  $(\mathbb F,\mathbb P)$-martingales and  \[M_t^{1,2}(u_1,u_2)=\frac{L_t^{1,2}(u_1,u_2)}{\alpha_t(u_1,u_2)},\quad t\geq\max(u_1,u_2).\]
In a recursive way, for $\{i,j\}=\{1,2\}$, let $(L^i_t(u_i),t\geq u_i)$ be $(\mathbb F,\mathbb P)$-martingales and \[M_t^i(u_i)=\frac{L^i_t(u_i)+\int_{u_i}^tM_{u_j}^{1,2}(u_1,u_2)
\alpha_{u_j}(u_1,u_2)du_j}{\int_t^\infty\alpha_t(u_1,u_2)du_j},\quad t\geq u_i.\]
Finally, let $(L^0_t,t\geq 0)$ be a $(\mathbb F,\mathbb P)$-martingale and \[M_t^0=\frac{L^0_t+\underset{\scriptscriptstyle{\{i,j\}=\{1,2\}}}{\sum}\int_0^tM_{u_i}^i(u_i)\int_{u_i}^\infty\alpha_{u_i}(u_1,u_2)du_jdu_i}{\int_t^\infty\alpha_t(u_1,u_2)du_1du_2},\quad t\geq 0.\]
Then the process $(M_t(\tau_1,\tau_2), t\geq 0)$ constructed as above is a $\mathbb G$-martingale.
\end{Exe}

\end{document}